\documentclass[a4paper]{IEEEtran}
\usepackage{amsmath}
\usepackage{amssymb}
\usepackage{graphicx}
\usepackage{subfig}
\usepackage{float}
\usepackage{tikz}
\usepackage{pgfplots}
\pgfplotsset{compat=1.14}
\usetikzlibrary{plotmarks}
\usetikzlibrary{datavisualization} 
\usetikzlibrary{patterns}
\usepackage{caption}
\usepackage{scalerel}
\usepackage[colorinlistoftodos]{todonotes}
\usepackage[noadjust]{cite}
\usepackage{hyperref}

\tikzset{>=latex}
\begin{document}

\title{Volterra-assisted Optical Phase Conjugation: a Hybrid Optical-Digital Scheme For Fiber Nonlinearity Compensation}

\author{Gabriel~Saavedra,~\IEEEmembership{Member,~IEEE,}
 		Gabriele~Liga,~\IEEEmembership{Member,~IEEE,} and~Polina~Bayvel~\IEEEmembership{Fellow,~IEEE,}~\IEEEmembership{Fellow,~OSA}
 \thanks{G. Saavedra, G. Liga, P. Bayvel are with the Optical Networks Group, Department of Electronic and Electrical Engineering, UCL, Torrington Place, London WC1E 7JE, UK. e-mail: (uceegsa@ucl.ac.uk).}
 \thanks{This work was supported by Becas Chile, UK EPSRC UNLOC (EP/J017582/1) and INSIGHT projects and continues under the TRANSNET Programme Grant (EP/R035342/1).}
}

\maketitle

\begin{abstract}
Mitigation of optical fiber nonlinearity is an active research field in the area of optical communications, due to the resulting marked improvement in transmission performance.
Following the resurgence of optical coherent detection, digital nonlinearity compensation (NLC) schemes such as digital backpropagation (DBP) and Volterra equalization have received much attention. 
Alternatively, optical NLC, and specifically optical phase conjugation (OPC), has been proposed to relax the digital signal processing complexity. 

In this work, a novel hybrid optical-digital NLC scheme combining OPC and a Volterra equalizer is proposed, termed Volterra-Assisted OPC (VAO). It has a twofold advantage: it overcomes the OPC limitation in asymmetric links and substantially enhances the performance of Volterra equalizers. The proposed scheme is shown to outperform both OPC and Volterra equalization alone by up to 4.2 dB in a 1000 km EDFA-amplified fiber link. Moreover, VAO is also demonstrated to be very robust when applied to long-transmission distances, with a 2.5 dB gain over OPC-only systems at 3000 km.       

VAO combines the advantages of both optical and digital NLC offering a promising trade-off between performance and complexity for future high-speed optical communication systems.       

\end{abstract}


\IEEEpeerreviewmaketitle

\section{Introduction}
\label{sec:introduction}
\IEEEPARstart{F}{ibre} nonlinearity represents a major bottleneck in increasing transmission rates of optical communication systems \cite{Essiambre2010,Agrell2016}.
As a result, finding a high-performance, low-complexity nonlinearity compensation (NLC) technique is a central challenge in the design of next-generation fibre systems. There are two main approaches to fiber nonlinearity mitigation, namely i) digital NLC techniques; and ii) optical NLC techniques. 
 
Digital NLC is based on digital signal processing (DSP) (either at the transmitter or at the receiver) to mitigate nonlinear distortions accumulated during signal propagation. Digital backpropagation (DBP) is indisputably the most widespread among these digital techniques \cite{Li2008,Ip2008}. DBP aims to numerically reverse the Manakov equation, which describes the signal evolution in an optical fiber, through the well-known split-step Fourier method (SSFM). The SSFM algorithm is recursive and in general complex, as a minimum number of iterations (or \emph{steps}), is required to achieve a given target performance. Such a minimum required number was shown to rapidly increase as a function of the NLC bandwidth \cite{Mateo2010, Liga2014}.

An alternative zero-forcing equalization method to DBP is based on the Volterra series expansion of the solution of the Manakov equation \cite{Peddanarappagari1997}. Volterra-based equalizers were first introduced for NLC in coherent optical communications in \cite{Gao2010, Pan2011}, initially through a time-domain approach, and subsequently \cite{Guiomar2011,Guiomar2012,Guiomar2013,Cartledge2017} in the frequency domain. 
A Volterra-series frequency-domain equalizer (VSFE) consists in reconstructing the transmitted optical field by using a Volterra series expansion of the Manakov equation (in the \emph{backward direction}). This expansion, first derived in \cite{Peddanarappagari1997}, is usually truncated to the third-order term for complexity reasons \cite{Guiomar2011}. This leads to suboptimal NLC performance, as nonlinear distortions are only approximately compensated for at the receiver. Previous works have shown a potential advantange in complexity for VSFE over DBP \cite{Guiomar2013,Liu2012,Bakhshali2016}, especially due of the highly parallel structure of the VSFE scheme.  

Alternative to digital NLC, optical NLC techniques are based on the idea of using the physics of fiber propagation itself to compensate for, or prevent detrimental nonlinear effects on the received signal. These techniques include optical phase conjugation (OPC) \cite{Yariv79, Fisher83, Watanabe96,Lorattanasane97} and phase-conjugated twin wave (PCTW) transmission \cite{XLiu2012}. Mid-link OPC consists in performing a conjugation of the optical field at the mid-point of the transmission link, such that nonlinear effects produced in the first part of the link are cancelled by the propagation effects in the second half.
Mid-link OPC can jointly compensate for chromatic dispersion and nonlinear effects within the transmission link \cite{Lorattanasane97,Yariv79}, with the big advantage, compared to digital NLC schemes, of compensating for inter-channel effects without requiring joint detection/processing of multiple channels.  
As such a compensation takes place in the optical domain, NLC can be performed jointly over very large optical bandwidths, e.g., in \cite{Umeki2016} NLC was achieved over 92 channels using OPC for long-haul transmission systems . Performing digital NLC over such bandwidths would be infeasible due to the high computational complexity. In addition, OPC can notionally provide nonlinearity suppression without sacrificing any of the available transmission degrees of freedom.  
A major shortcoming of OPC is the stringent requirement of a symmetric link power profile. Optical links with symmetric power profile at the mid-link point can only be designed using distributed amplification schemes. Moreover, even with Raman amplifiers, the power profile symmetry condition requires complex amplifier configurations \cite{Rosa2015,Ellis2014} that can be hard to achieve in practice. On the other hand, performing OPC in links with a certain degree of power profile asymmetry can result in a significantly reduced ability to reduce fibre nonlinear effects \cite{Saavedra2018,Watanabe96}. In addition to this, OPC devices are never ideal as they typically include an optical amplifier to overcome negative conversion efficiencies and add amplified spontaneous emission (ASE) noise in the process. This leads to a performance penalty, which further reduces their nonlinearity mitigation capabilities \cite{Saavedra2018}.

In \cite{Liga2018}, a novel approach to nonlinearity compensation was proposed that combined OPC and Volterra equalization, termed Volterra-assisted OPC (VAO). In this paper, we extend the results through a detailed analysis of the advantages of VAO over each of its constituent techniques. 
The proposed approach enhances the gain of OPC in links with a low degree of power profile symmetry, such as EDFA-amplified links, through the use of VSFE. Moreover, it applies OPC as a way to boost the NLC performance of VSFE in large NLC bandwidth scenarios.  

Hybrid optical NLC schemes were previously studied, e.g. in \cite{Cartledge2016}, 
with joint operation of OPC and DBP. However, this combination was designed to relax the link placement of the OPC device, rather than a way to enhance NLC performance. The technique proposed in this work is shown to markedly outperform either OPC or VSFE schemes when individually used in EDFA-amplified multi-span transmission systems, over large NLC bandwidths.
\begin{figure*}[t]
\centering
\subfloat[]{
\includegraphics[width=0.95\columnwidth,keepaspectratio]{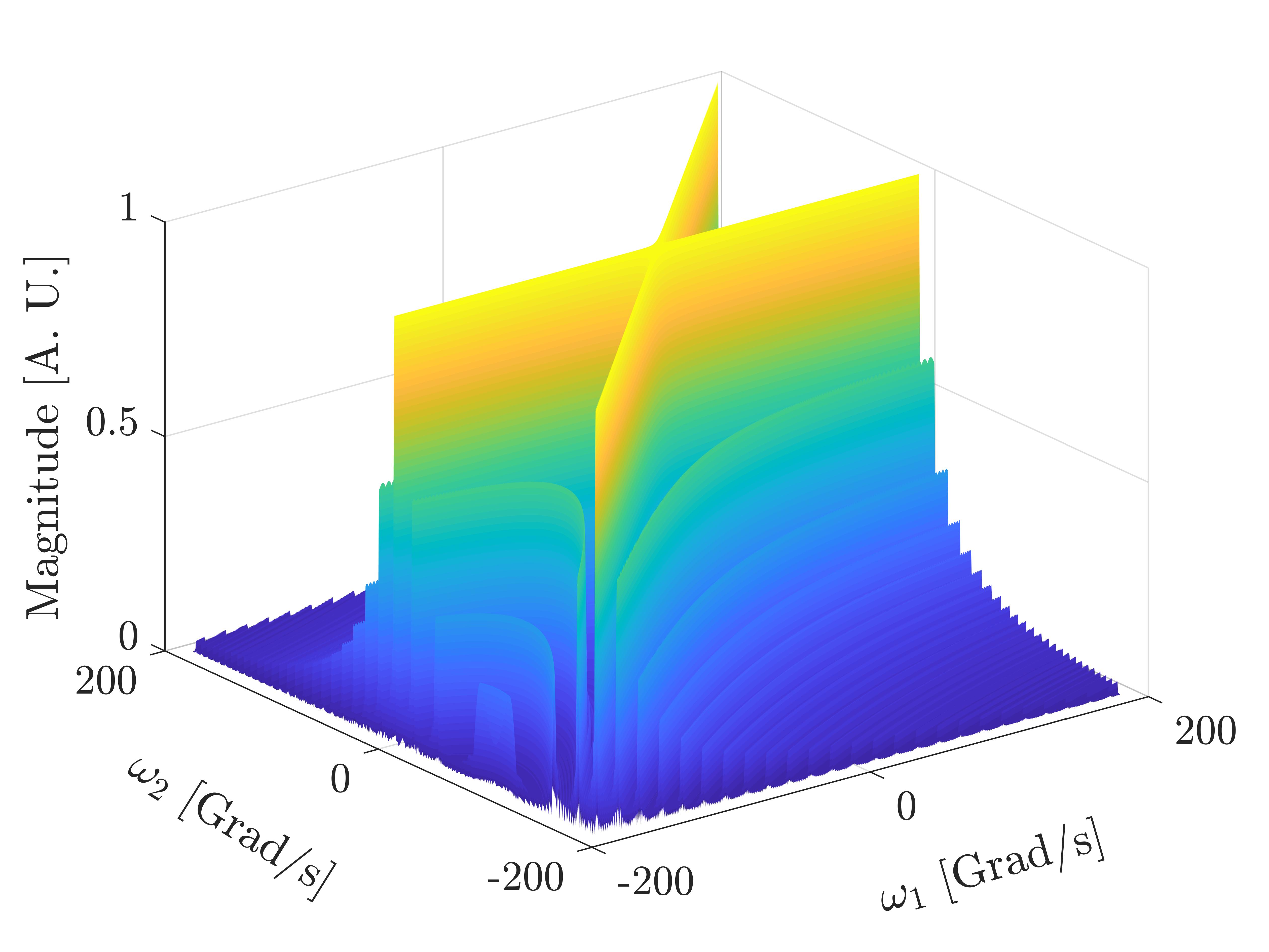}\label{subfig:kernel_a}}
\subfloat[]{
\includegraphics[width=0.95\columnwidth,keepaspectratio]{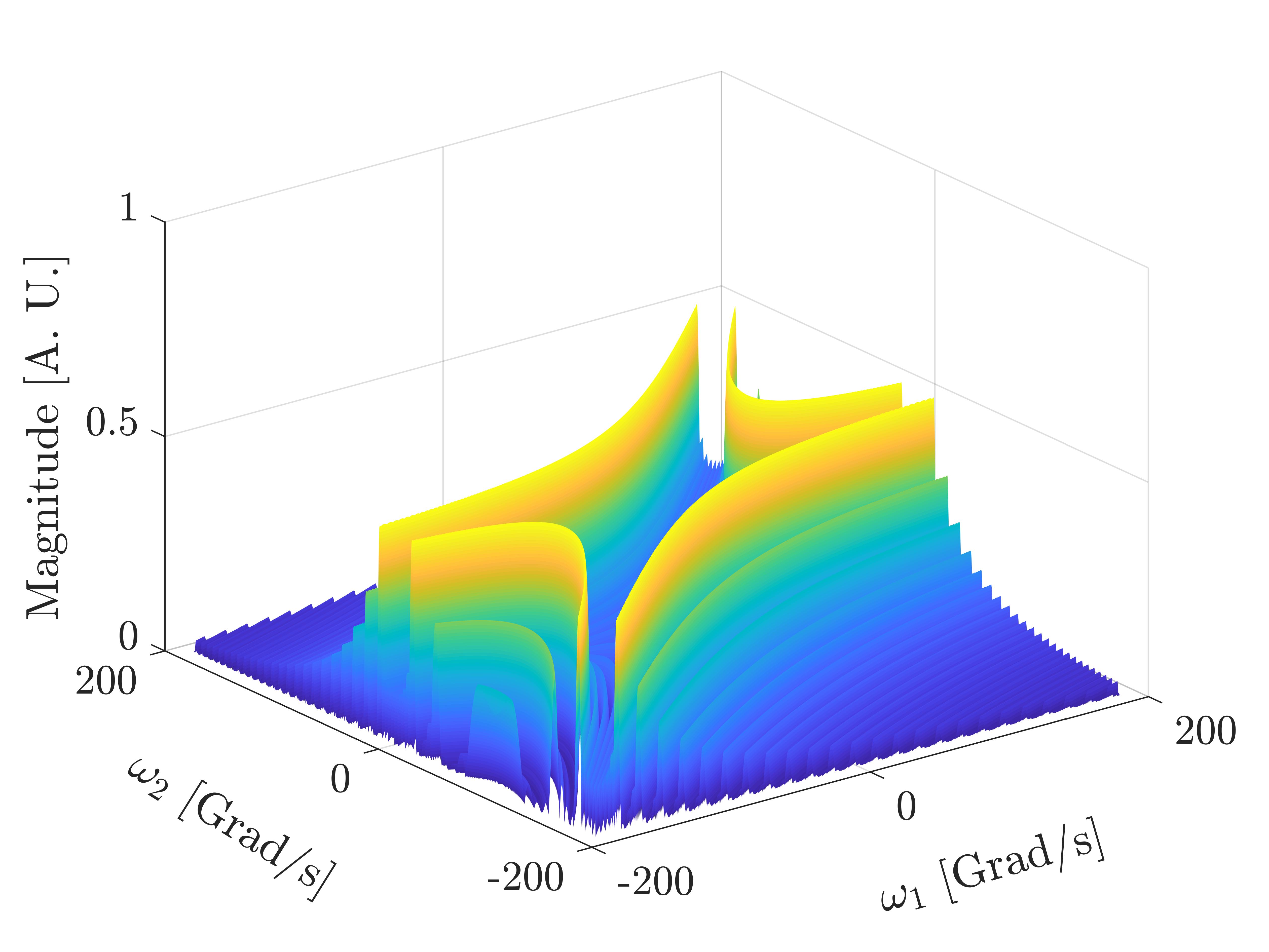}\label{subfig:kernel_b}}
\caption{Third-order nonlinearity kernels for a 10 $\times$ 100 km SSMF, EDFA-amplified link \protect\subref{subfig:kernel_a} without OPC, and \protect\subref{subfig:kernel_b} with OPC.}
\label{fig:VolterraKernels}
\end{figure*}

\section{Volterra-assisted OPC}
\label{sec:VAO}
To analyze the effect of nonlinearities generated in an optical fiber transmission system, a possible approach is to use a frequency domain Volterra series expansion. 
Usually, for optical fibre propagation, a truncated Volterra series up to the third-order term is adopted, due to the numerical complexity in calculating the higher order terms of the series. The third-order term coincides with the first-order regular-perturbation solution to the Manakov equation (see \cite{Vannucci2002}) and is a powerful analytical tool to explore the characteristics of the nonlinear interference in the optical fiber channel \cite{Peddanarappagari1997, Taghavi2006}. For multi-span, EDFA-amplified, dispersion-uncompensated transmission links, the third-order Volterra series term is given by\footnote{In this work, functions are denoted by uppercase letters $G(\cdot)$, boldface uppercase letters (e.g. $\boldsymbol{\Xi}$) denote matrices, whereas lowercase letters with either 2 or 3 sub-index (e.g. $a_{ilk}$) denote elements of a 2D/3D tensor.}: 
\begin{align}
\begin{split}
&A_{3\scaleto{X}{4pt}}(\omega)=j\gamma\frac{8}{9}\iint \left[\mathcal{S}_{\scaleto{XXX}{4pt}}(\omega,\omega_1,\omega_2)\right.\\
&\left.+\mathcal{S}_{\scaleto{YYX}{4pt}}(\omega,\omega_1,\omega_2)\right]\Xi\left(N_s,\omega,\omega_1,\omega_2\right)\\
&\times F(\omega,\omega_1,\omega_2)  d\omega_2 d\omega_1,
\end{split}
\label{eq:VolterraInt}
\end{align}
where $\gamma$ is the nonlinearity coefficient of the fiber,
\begin{align}
\begin{split}
\mathcal{S}_{\scaleto{XXX}{4pt}}(\omega,\omega_1,\omega_2)\triangleq S^{*}_{\scaleto{X}{4pt}}(\omega_1)S_{\scaleto{X}{4pt}}(\omega_2)S_{\scaleto{X}{4pt}}(\omega+\omega_1-\omega_2),
\\
\mathcal{S}_{\scaleto{YYX}{4pt}}(\omega,\omega_1,\omega_2)\triangleq S^*_{\scaleto{Y}{4pt}}(\omega_1)S_{\scaleto{Y}{4pt}}(\omega_2)S_{\scaleto{X}{4pt}}(\omega+\omega_1-\omega_2),
\end{split}
\label{eq:SignalKernels}
\end{align}
are the \emph{signal kernels}, and $S_{\scaleto{X}{3pt}}(\omega)$, $S_{\scaleto{Y}{3pt}}(\omega)$ are the transmitted signal spectra over two orthogonal polarizations $X$ and $Y$, respectively. $\Xi\left(N_s,\omega,\omega_1,\omega_2\right)$
is the \emph{phased-array} term given by 
\begin{equation}
\Xi\left(N_s,\omega,\omega_1,\omega_2\right)=\sum_{n=1}^{N_s}e^{-j\beta_2 \Delta\Omega(n-1)L_s},
\label{eq:pharray}
\end{equation}
with $\Delta\Omega\triangleq(\omega-\omega_2)(\omega_1-\omega_2)$, and where $\beta_2$, $L_s$ ,and $N_s$  are the group velocity dispersion, the fiber span length, and the total number of spans, respectively. The factor $F(\omega,\omega_1,\omega_2)$ represents the \emph{four-wave mixing efficiency} over one span, given by
\begin{equation}
F(\omega,\omega_1,\omega_2) = \frac{1-e^{-\alpha L_s} e^{j\beta_2\Delta\Omega L_s}}{j\beta_2\Delta\Omega-\alpha},
\label{eq:F1span}
\end{equation}
where $\alpha$ is the fiber attenuation coefficient. The function of $\omega_1$ and $\omega_2$ that multiplies the signal kernel (in this case $F(\omega,\omega_1,\omega_2)\cdot \Xi\left(N_s,\omega,\omega_1,\omega_2\right) $) is usually referred to \emph{Volterra third-order kernel}. 

When an OPC device is inserted in the transmission link with EDFA amplification, only part of the nonlinear distortion accumulated in the first part of the link is effectively cancelled by the OPC. Such a residual nonlinear term can be described using a third-order Volterra-series term. Similar to the case without OPC, the Volterra kernel can be obtained using a regular perturbation approach for this specific system configuration (see Appendix \ref{appendix}). For a transmission link using mid-link OPC, the first and third-order Volterra series terms in frequency domain, are given by\footnote{Due to the symmetry of the Manakov equation, $A_{1\scaleto{Y}{4pt}}$ and $A_{3\scaleto{Y}{4pt}}$ can be obtained by simply swapping sub-indexes $X$ and $Y$ in \eqref{eq:VolterraInt_OPC} and \eqref{eq:SignalKernels}.} 
\begin{align}
\begin{split}
&A_{1\scaleto{X}{4pt}}(\omega)=S^*_{\scaleto{X}{4pt}}(-\omega),
\end{split}
\label{eq:VolterraFirstOrder}
\end{align}

\begin{align}
\begin{split}
&A_{3\scaleto{X}{4pt}}(\omega)=j\gamma\frac{8}{9}\iint \left[\mathcal{S}_{\scaleto{XXX}{4pt}}^*(-\omega,-\omega_1,-\omega_2)\right.\\
&\left.+\mathcal{S}_{\scaleto{YYX}{4pt}}^*(-\omega,-\omega_1,-\omega_2)\right]\Xi^*\left(\frac{N_s}{2},\omega,\omega_1,\omega_2\right)\\
&\times G(\omega,\omega_1,\omega_2)  d\omega_2 d\omega_1.
\end{split}
\label{eq:VolterraInt_OPC}
\end{align}

The first-order Volterra series term simply corresponds to the conjugated version of the transmitted signal, as chromatic dispersion compensation is applied in-line by the mid-link OPC. Moreover, it is interesting to notice how the phased-array in the third-order term (Eq.~\eqref{eq:VolterraInt_OPC}) only accounts for the accumulation of nonlinear effects over half of the link ($\frac{N_s}{2}$ fiber spans for a link of $N_s$ spans). Finally, $G(\omega,\omega_1,\omega_2)$ is the characteristic nonlinear kernel for systems with OPC and is given by\ (see Appendix \ref{appendix})
\begin{align}
\begin{split}
G(\omega,\omega_1,\omega_2)&\triangleq\frac{(e^{-j\beta_2\Delta\Omega L_s}e^{-\alpha L_s}-1)(\alpha-j\beta_2\Delta\Omega)}{\alpha^2+\beta_2^2\Delta\Omega^2}
\\&+\frac{(e^{-j\beta_2\Delta\Omega L_s}-e^{-\alpha L_s})(\alpha+j\beta_2\Delta\Omega)}{\alpha^2+\beta_2^2\Delta\Omega^2}.
\end{split}
\label{eq:OPC_kernel1}
\end{align}

Fig.~\ref{fig:VolterraKernels} shows the absolute value of the normalized frequency-domain third-order Volterra kernels in Eqs.~\eqref{eq:VolterraInt} (Fig.~\ref{subfig:kernel_a}) and \eqref{eq:OPC_kernel1} (Fig.~\ref{subfig:kernel_b}) as a function of the 2 angular frequency variables $\omega_1$ and $\omega_2$  for $\omega=0$, for an EDFA-amplified transmission link of 10x100 km spans of standard single mode fiber (SSMF). The OPC channel kernel in Fig.~\ref{subfig:kernel_b} is here normalised with respect to the maximum value of the kernel in \eqref{eq:VolterraInt}. As expected, the kernel of the channel without OPC (Fig.~\ref{subfig:kernel_a}) shows a maximum coupling for $\omega_1=\omega_2=0$ and $\omega_1=\omega_2$. A decrease in the coupling strength is instead observed as $|\omega_1- \omega_2|\gg 0$ and $\omega_2\neq 0$, due to the walk-off effect induced by fiber dispersion. The oscillations observed in the kernel arise from the coherent accumulation of nonlinearities at every span and are described by the phased-array factor.
 
From \ref{subfig:kernel_b} it can be seen that OPC produces a kernel with a characteristic \emph{dip} around the $\left(0, 0 \right)$ frequency and along the $\omega_1=\omega_2$ and $\omega_2=0$ lines. This indicates a strong reduction of the nonlinear coupling in the low-frequency $\left(\omega_1 , \omega_2\right)$--plane. On the other hand, higher frequency coupling is still present due to the power profile asymmetry\footnote{Ideal mid-link OPC would have zero nonlinear coupling everywhere in the $\left(\omega_1 , \omega_2\right)$--plane.}. As it can be observed from Eq.~\eqref{eq:VolterraInt_OPC}, where the phased-array is evaluated only over $N_s/2$ spans, OPC reduces the maximum magnitude of the nonlinear kernel by a factor of 2. 
The residual nonlinear interactions at the receiver, due to the lack of symmetry in EDFA-based links, limit the performance of OPC as an NLC method. Similar conclusions were previously derived using a different mathematical approach, such as the four-wave mixing product analysis presented in \cite{Pechenkin2010,Al-Khateeb2018} and follow-up publications \cite{Al-Khateeb:18}. However, no mitigation strategy specifically tailored to overcome this limitation has been proposed to date.

Eqs.~\eqref{eq:VolterraInt_OPC} and \eqref{eq:OPC_kernel1} specify the design of a VSFE tailored to compensate for the residual nonlinearity in an EDFA-amplified link using OPC. A VSFE aims to reconstruct the transmitted field up to the third-order Volterra series term, through a frequency-domain processing. This reconstruction can be done using: i) a fully non-recursive equalizer, that calculates the third-order Volterra kernel of the entire link in a single step (multispan VSFE) \cite{Peddanarappagari1997},\cite{Guiomar2013}[Sec.~2]; ii) a recursive, per-step, equalizer \cite{Guiomar2011,Guiomar2012}. Although the recursive method partly mitigates the so-called \emph{energy divergence} problem, which leads to extremely inaccurate field reconstructions as the transmitted power increases \cite{Xu2002}, it also requires $N_s$ times more calculations, for a fixed processed sequence length.

To take advantage of the action of the OPC, the VAO system proposed in this work naturally requires a non-recursive VSFE approach based on the kernel in \eqref{eq:VolterraInt_OPC} (as opposed to a per-span fiber kernel). This is due to the fact that the equalizer is supposed to benefit from the properties of the end-to-end kernel of the OPC channel, as opposed to the per-span kernel. The reason for that is twofold and it is fully justified from the results in this paper: i) OPC reduces the memory of the system as chromatic dispersion is compensated in a distributed manner along the link, thus allowing for a decreased interplay between nonlinearity and chromatic dispersion; ii) the reduction of nonlinearity induced by the OPC leads to a more accurate reconstruction of transmitted field using third-order Volterra equalizers.

\section{VSFE and VAO implementation}
\label{sec:VAOimplementation}
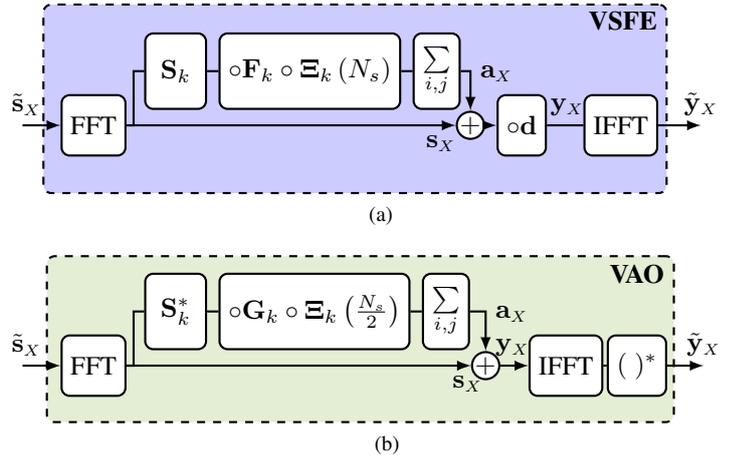
\begin{figure}
\subfloat[]{
\definecolor{mycolor5}{rgb}{0.46600,0.67400,0.18800}%

\hspace{-.55cm}
\begin{tikzpicture}
[plain/.style={align=center,execute at begin node=\setlength{\baselineskip}{2.5ex}}]

\node[plain]  at (-200pt,0pt) (input){};

\node[thick,rounded corners=3pt,fill=blue,fill opacity=0.2,dashed, minimum height=2.5cm, minimum width=8.2cm, draw,label={[label distance=-1.2cm]15:\textbf{VSFE}}] at (-72pt,10pt) (VSFE) {};

\node[thick,rounded corners=3pt, minimum height=0.8cm, minimum width=.6cm, draw,fill=white, right=0.5cm of input] (fft) {FFT};

\node[thick,rounded corners=3pt, minimum height=1cm, minimum width=.8cm, draw, fill=white, above right=0.08cm and 1.6cm of input] (sig_kernel) {$\mathbf{S}_k$};

\node[thick,rounded corners=3pt, minimum height=1cm, minimum width=1.3cm, draw, fill=white, right=0.15cm of sig_kernel] (hadamard) {$\circ \mathbf{F}_{k}\circ \boldsymbol{\Xi}_{k}\left(N_s\right)$};

\draw[thick] (fft)--++(15pt,0pt) node[](a){}--(a|-sig_kernel)--(sig_kernel)--(hadamard);

\node[thick,rounded corners=3pt, draw, right=0.15cm of hadamard,fill=white,minimum width=0.6cm,minimum height=1cm] (sumsum) {$\sum\limits_{i,j}$};


\node[thick,rounded corners=3pt, draw, right=5.7cm of input,fill=white,circle,inner sep=0pt] (sum) {$+$};


\node[thick,rounded corners=3pt, minimum height=0.8cm, minimum width=0.5cm, draw,fill=white, right=0.15cm of sum] (disp_comp) {$\circ  \mathbf{d}$};

\node[thick,rounded corners=3pt, minimum height=0.8cm, minimum width=.6cm, draw,fill=white,right=0.48cm of disp_comp] (ifft) {IFFT};


\draw[thick] (hadamard)--(sumsum);
\draw[thick] (disp_comp)--++(17pt,0pt) node[above]{$\mathbf{y}_{\scaleto{X}{4pt}}$} --(ifft);
\draw[thick,->] (input)--++(5pt,0pt) node[above]{$\tilde{\mathbf{s}}_{\scaleto{X}{4pt}}$}--(fft);
\draw[thick,->] (fft)--++(130pt,0pt) node[below]{$\mathbf{s}_{\scaleto{X}{4pt}}$} --(sum);

\draw[thick,->] (sumsum)--(sumsum-|sum)--++(0pt,-1pt) node[right] () {$\mathbf{a}_{\scaleto{X}{4pt}}$}--(sum);
\draw[thick,->] (sum)--(disp_comp);
\draw[thick,->] (ifft)--++(30pt,0pt) node[above] {$\tilde{\mathbf{y}}_{\scaleto{X}{4pt}}$};

\end{tikzpicture} \\
\label{subfig:VSFE}
} \\
\subfloat[]{
\definecolor{mycolor5}{rgb}{0.46600,0.67400,0.18800}%
\hspace{-.55cm}
\begin{tikzpicture}[plain/.style={align=center,execute at begin node=\setlength{\baselineskip}{2.5ex}}]
\node[plain] at (-200pt,0pt) (input){};

\node[thick,rounded corners=3pt,fill=mycolor5,fill opacity=0.2,dashed, minimum height=2.2cm, minimum width=8.25cm, draw,label={[label distance=-1cm]11:\textbf{VAO}}] at (-70pt,10pt) (VAO) {};

\node[thick,rounded corners=3pt, minimum height=0.8cm, minimum width=.6cm, draw,fill=white, right=0.5cm of input] (fft) {FFT};

\node[thick,rounded corners=3pt, minimum height=1cm, minimum width=.8cm, draw, fill=white, above right=0.08cm and 1.6cm of input] (sig_kernel) {$\mathbf{S}^*_k$};

\node[thick,rounded corners=3pt, minimum height=1cm, minimum width=1cm, draw, fill=white, right=0.15cm of sig_kernel] (hadamard) {$\circ \mathbf{G}_{k}\circ \boldsymbol{\Xi}_{k}\left(\frac{N_s}{2}\right)$};

\draw[thick] (fft)--++(15pt,0pt) node[](a){}--(a|-sig_kernel)--(sig_kernel)--(hadamard);

\node[thick,rounded corners=3pt, draw, right=0.15cm of hadamard,fill=white,minimum width=0.5cm,minimum height=1cm] (sumsum) {$\sum\limits_{i,j}$};

\node[thick,rounded corners=3pt, draw, right=5.9cm of input,fill=white,circle,inner sep=0pt] (sum) {$+$};

\node[thick,rounded corners=3pt, draw, right=.37cm of sum,fill=white,minimum height=0.8cm, minimum width=.6cm] (ifft) {IFFT};

\node[thick,rounded corners=3pt, draw, right=1.4cm of sum,fill=white,minimum height=0.8cm] (conj) {$(\;)^*$};




\draw[thick] (hadamard)--(sumsum);
\draw[thick,->] (input)--++(5pt,0pt) node[above]{$\tilde{\mathbf{s}}_{\scaleto{X}{4pt}}$}--(fft);
\draw[thick,->] (sum)--++(10pt,0pt) node[above]{$\mathbf{y}_{\scaleto{X}{4pt}}$}--(ifft);
\draw[thick,->] (sumsum)--(sumsum-|sum)--++(0pt,-1pt) node[right] () {$\mathbf{a}_{\scaleto{X}{4pt}}$}--(sum);
\draw[thick] (ifft)--(conj);
\draw[thick,->] (conj)--++(25pt,0pt) node[above]{$\tilde{\mathbf{y}}_{\scaleto{X}{4pt}}$};
\draw[thick,->] (fft)--++(140pt,0pt) node[below]{$\mathbf{s}_{\scaleto{X}{4pt}}$} --(sum);

\end{tikzpicture}
\label{subfig:VAO}
}
\caption{Schematic diagram of the two Volterra equalizers compared in this work: a \protect\subref{subfig:VSFE} conventional VSFE, and \protect\subref{subfig:VAO} the equalizer in the proposed VAO scheme.}
\label{fig:DSP}
\end{figure}

Fig.~\ref{fig:DSP} shows the difference between a conventional (non-recursive) VSFE equalizer and the Volterra equalizer in VAO. In both equalizers, $N$ samples of the received time-signal  (on each polarization) $\tilde{\mathbf{s}}_{\scaleto{X}{4pt}}\triangleq \{\tilde{s}_{\scaleto{X}{4pt},k}\}_{k=0}^{N-1}$ over a time window $T$, are transformed into the frequency-domain vector $\mathbf{s}_{\scaleto{X}{4pt}}\triangleq \{s_{\scaleto{X}{4pt},k}\}_{k=0}^{N-1}$, where 
\begin{equation}
s_{\scaleto{X}{4pt},k}=N\cdot\Delta\omega\cdot S_{\scaleto{X}{4pt}}\left(k\Delta\omega\right) ,\;\; \quad k=-\frac{N}{2},-\frac{N}{2}+1,...,\frac{N}{2}-1 
\end{equation} and $\Delta\omega=\frac{1}{T}$ represents the sampling resolution in frequency-domain. 
The signal kernel matrix $\mathcal{S}_k\triangleq (\mathbf{\sigma}_{ij}=s^*_i s_j s_{k+i-l})$ (or $\mathcal{S}^{*}_k$ in the case of VAO) is then computed and an Hadamard multiplication by the respective channel kernel matrix is performed\footnote{The Hadamard multiplication, denoted by the $\circ$ symbol in Fig.~\ref{fig:DSP},  is a simple element-wise matrix multiplication.}. 
The kernel matrices $\mathbf{F}_k \circ \boldsymbol{\Xi}_k(N_s)$, and $\boldsymbol{G}_k \circ \boldsymbol{\Xi}_k\left( \frac{N_s}{2} \right)$, for, respectively, the VSFE and VAO cases, are defined as 
\begin{align}
\begin{split}
\mathbf{F}_k&\triangleq  (f_{ilk}=F^{\prime}\left(k\Delta\omega,i\Delta\omega,l\Delta\omega\right)),   \\
\boldsymbol{\Xi}_k&\triangleq (\xi_{ilk}=\Xi^{\prime}\left(k\Delta\omega,i\Delta\omega,l\Delta\omega^{\prime}\right)),   \\
\boldsymbol{G}_k&\triangleq (g_{ilk}=G^{\prime}\left(k\Delta\omega,i\Delta\omega,l\Delta\omega\right)),\\
&\forall \quad k,i,l=-\frac{N}{2},-\frac{N}{2}+1,...,\frac{N}{2}-1 
\label{eq:ChannelKernels}
\end{split}
\end{align}
where $F^{\prime}$, $\Xi^{\prime}$, and $G^{\prime}$ are the backward direction counterparts of the functions $F$, $\Xi$, and $G$ (see Appendix \ref{appendix}). 

Finally, both equalizers compute the reconstructed field (in frequency-domain) which, in the VAO case, is

\begin{align}
\begin{split}
y_{\scaleto{X}{4pt},k}&=s_{\scaleto{X}{4pt},k}+a_{\scaleto{X}{4pt},k},\\ 
k&=-\frac{N}{2},-\frac{N}{2}+1,...,\frac{N}{2}-1. 
\end{split}
\label{eq:VSFEimplementation}
\end{align}

The sequence $a_{\scaleto{X}{4pt},k}$ is the sampled version of the third-order Volterra series terms in Eqs. \eqref{eq:VolterraInt} and \eqref{eq:VolterraInt_OPC} and is given by
\begin{align}
\begin{split}
a_{\scaleto{X}{4pt},k}=F_s &A_{3\scaleto{X}{4pt}}\left(k\Delta\omega\right)\\=\frac{1}{N ^2}\sum_{i=0}^{N-1}\sum_{l=\max(0, k+i-N+1)}^{\min(k+i, N)}&\xi_{ilk}g_{ilk}\left[s_{\scaleto{X}{4pt},i}s^*_{\scaleto{X}{4pt},l}s^*_{\scaleto{X}{4pt},k+i-l}\right.\\ \left. +s_{\scaleto{Y}{4pt},i}s^*_{\scaleto{Y}{4pt},l}s^*_{\scaleto{X}{4pt},k+i-l}\right].
\label{eq:VSFEdoublesum}
\end{split}
\end{align}

In the VSFE case, the chromatic dispersion accumulated in the link is then removed through the multiplication by $\mathbf{d}\triangleq \left\lbrace\exp\left(-j\frac{\beta_2}{2}k^2\Delta\omega^2\right) \right\rbrace_{k=-\frac{N}{2}}^{\frac{N}{2}-1}$. Finally the sequence is transformed into the time-domain and, in the VAO case, and conjugated back to undo the effect of the OPC module (see Appendix \ref{appendix}).  

\label{sec:numerical_study}

In this work, the proposed VAO method was numerically assessed in comparison with two conventional VSFE implementations: single-step and recursive VSFE.
The single-step version of the VSFE was implemented using the schematic diagram in Fig.~\ref{subfig:kernel_a}. The iterative VSFE was implemented using a per-span equalizer, where only the $\mathbf{F}_k$ kernel is used to calculate $a_{\scaleto{X}{4pt},k}$ at each span (see \cite{Guiomar2012}). At the end of each iteration, the signal is transformed back into time-domain to include the modification to the output signal presented in \cite{Xu2002}. Modifying the Volterra output is crucial to account for the previously mentioned energy divergence effect (see Sec.~\ref{sec:VAO}). 

The computational complexity of a VSFE, scales between $O(N)$ and $O(N^2)$ per sample, depending on the specific implementation adopted \cite{Guiomar2011,Guiomar2012}. For continuous data transmission, this effectively limits the window size over which the VSFE can be sequentially applied. Conversely, a minimum window size must be guaranteed to allow for an absorbing window for the fiber channel memory effects to be factored out. This is done typically through an overlap and save approach, where a certain number of symbols are discarded from each side of the window, to account for the cyclic effects induced by the discrete frequency-domain operation. 
In this work, VSFE was applied sequentially over subsequent sequences of samples, whose size was varied (between 128 and 1024 symbol periods) depending on the transmission distance. For each processed sequence, a certain amount of symbols were then discarded from each side of the window, to account for the above mentioned cyclic effects. Despite the increased complexity per sample (see e.g. \cite{Bakhshali2016}), the discarding operation considerably benefits the equalizer performance. As discussed in the following sections, the performance metrics analysed in this work (NLI suppression and SNR) are closely related to the reduction of NLI distortion. This is affected by the window size used for the VSFE. In order to guarantee the best performance, the VSFE window size was set to 4 times (up to a maximum of 1024 symbols) the estimated channel memory. A rough estimate of the channel memory can be obtained using $M\approx |\beta_2|\cdot R_s \cdot B\cdot L$  where $R_s$ is the channel symbol rate, $B$ is the entire transmitted bandwidth, and $L$ is the total transmitted distance. Each sub-sequence was sequentially processed by the VSFE, and then demodulated. The boundary symbols of the demodulated subsequence were then increasingly discarded until the  \emph{performance metric} of choice converged within a desired tolerance. Multiple sub-sequences were then processed in order to accumulate enough received symbols and to guarantee the accuracy of \emph{performance metric} calculation. 

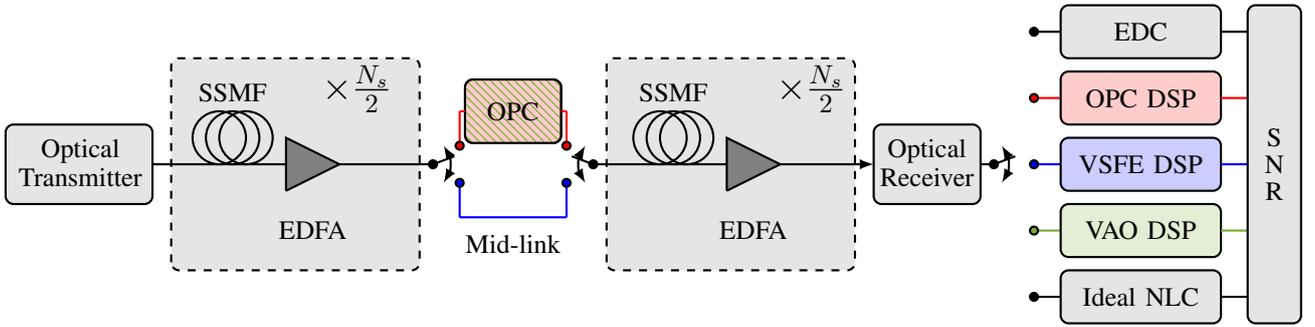
\begin{figure*}
\centering
\definecolor{mycolor5}{rgb}{0.46600,0.67400,0.18800}%
\begin{tikzpicture}
[plain/.style={align=center,execute at begin node=\setlength{\baselineskip}{2.5ex}}]

\draw[thick,rounded corners=3pt,fill=gray,fill opacity=0.2] (-225pt,-15pt) rectangle (-170pt,15pt);
\node[plain] at (-197pt,5pt) {Optical};
\node[plain] at (-197pt,-5pt) {Transmitter};
\draw[thick,fill=gray,fill opacity=0.2,dashed,rounded corners=3pt] (-163pt,-40pt) rectangle (-70pt,40pt);
\node[plain,anchor=south west] at (-110pt,16pt) {\Large $\times \frac{N_s}{2}$};
\node[plain,anchor=south] at (-140pt,20pt) {SSMF};
\draw[thick,-] (-145pt,10pt) circle (10pt);
\draw[thick,-] (-140pt,10pt) circle (10pt);
\draw[thick,-] (-135pt,10pt) circle (10pt);
\draw[thick,-] (-170pt,0pt) -- (-65pt,0pt); 
\draw[thick,-,fill=black] (-65pt,0pt) circle (1.5pt);
\draw[thick,-] (-65pt,0pt) -- (-57pt,3pt); 
\draw [thick, <->] (-61pt,-7pt) to [bend right] (-61pt,7pt);
\draw[thick,-,fill=gray] (-120pt,-10pt) -- (-100pt,0pt) -- (-120pt,10pt) -- (-120pt,-10pt);
\node[plain] at (-110pt,-25pt) {EDFA};
\draw[thick,fill=red,fill opacity=0.2,rounded corners=3pt] (-53pt,8pt) rectangle (-17pt,32pt);
\draw[thick,rounded corners=3pt,pattern=north west lines, pattern color=mycolor5] (-53pt,8pt) rectangle (-17pt,32pt);
\node[plain] at (-35pt,20pt) {OPC};
\node[plain] at (-35pt,-30pt) {Mid-link};
\draw[thick,-,red] (-55pt,7pt) -- (-55pt,20pt);
\draw[thick,-,fill=red] (-55pt,7pt) circle (1.5pt);
\draw[thick,-,red] (-15pt,7pt) -- (-15pt,20pt);
\draw[thick,-,fill=red] (-15pt,7pt) circle (1.5pt);
\draw[thick,-,red] (-55pt,20pt) -- (-53pt,20pt);
\draw[thick,-,red] (-15pt,20pt) -- (-17pt,20pt);
\draw[thick,-,blue] (-55pt,-7pt) -- (-55pt,-20pt);
\draw[thick,-,fill=blue] (-55pt,-7pt) circle (1.5pt);
\draw[thick,-,blue] (-15pt,-7pt) -- (-15pt,-20pt);
\draw[thick,-,fill=blue] (-15pt,-7pt) circle (1.5pt);
\draw[thick,-,blue] (-55pt,-20pt) -- (-15pt,-20pt);
\draw[thick,-,fill=black] (-5pt,0pt) circle (1.5pt);
\draw[thick,-] (-5pt,0pt) -- (-13pt,3pt); 
\draw [thick, <->] (-8pt,-7pt) to [bend left] (-8pt,7pt);
\draw[thick,dashed,rounded corners=3pt,fill=gray,fill opacity=0.2] (0pt,-40pt) rectangle (93pt,40pt);
\node[plain,anchor=south west] at (60pt,16pt) {\Large $\times \frac{N_s}{2}$};
\draw[thick,->] (-5pt,0pt) -- (100pt,0pt); 
\draw[thick,-] (20pt,10pt) circle (10pt);
\draw[thick,-] (25pt,10pt) circle (10pt);
\draw[thick,-] (30pt,10pt) circle (10pt);
\draw[thick,-,fill=gray] (45pt,-10pt) -- (65pt,0pt) -- (45pt,10pt) -- (45pt,-10pt);
\node[plain] at (55pt,-25pt) {EDFA};
\node[plain,anchor=south] at (25pt,20pt) {SSMF};
\draw[thick,rounded corners=3pt,fill=gray,fill opacity=0.2] (100pt,-15pt) rectangle (140pt,15pt);
\node[plain] at (120pt,5pt) {Optical};
\node[plain] at (120pt,-5pt) {Receiver};
\draw[thick,-] (140pt,0pt) -- (145pt,0pt);
\draw[thick,-] (145pt,0pt) -- (153pt,3pt); 
\draw[thick,-,fill=black] (145pt,0pt) circle (1.5pt);
\draw [thick, <->] (148pt,-7pt) to [bend right] (148pt,7pt);
\draw[thick,-,blue] (160pt,0pt) -- (170pt,0pt);
\draw[thick,-,fill=blue] (160pt,0pt) circle (1.5pt);
\draw[thick,fill=none,rounded corners=3pt,fill=blue,fill opacity=0.2] (170pt,-10pt) rectangle (230pt,10pt);
\node[plain] at (200pt,0pt) {VSFE DSP};
\draw[thick,-,red] (160pt,25pt) -- (170pt,25pt);
\draw[thick,-,fill=red] (160pt,25pt) circle (1.5pt);
\draw[thick,rounded corners=3pt,fill=red,fill opacity=0.2] (170pt,15pt) rectangle (230pt,35pt);
\node[plain] at (200pt,25pt) {OPC DSP};
\draw[thick,-,mycolor5] (160pt,-25pt) -- (170pt,-25pt);
\draw[thick,-,fill=mycolor5] (160pt,-25pt) circle (1.5pt);
\draw[thick,rounded corners=3pt,fill=mycolor5,fill opacity=0.2] (170pt,-15pt) rectangle (230pt,-35pt);
\node[plain] at (200pt,-25pt) {VAO DSP};
\draw[thick,rounded corners=3pt,fill=gray,fill opacity=0.2] (170pt,40pt) rectangle (230pt,60pt);
\draw[thick,-] (160pt,50pt) -- (170pt,50pt);
\draw[thick,-,fill = black] (160pt,50pt) circle (1.5pt);
\node[plain] at (200pt,50pt) {EDC};
\draw[thick,rounded corners=3pt,fill=gray,fill opacity=0.2] (170pt,-40pt) rectangle (230pt,-60pt);
\draw[thick,-,black] (160pt,-50pt) -- (170pt,-50pt);
\draw[thick,-,fill=black] (160pt,-50pt) circle (1.5pt);
\node[plain] at (200pt,-50pt) {Ideal NLC};
\draw[thick,-,black] (230pt,50pt) -- (240pt,50pt);
\draw[thick,-,red] (230pt,25pt) -- (240pt,25pt);
\draw[thick,-,blue] (230pt,0pt) -- (240pt,0pt);
\draw[thick,-,mycolor5] (230pt,-25pt) -- (240pt,-25pt);
\draw[thick,-,black] (230pt,-50pt) -- (240pt,-50pt);
\draw[thick,fill=none,rounded corners=3pt,fill=gray,fill opacity=0.2] (240pt,-60pt) rectangle (260pt,60pt);
\node[plain] at (250pt,10pt) {S};
\node[plain] at (250pt,0pt) {N};
\node[plain] at (250pt,-10pt) {R};
\end{tikzpicture}
\caption{Systems compared in the numerical investigation in this work.}
\label{fig:setup}
\end{figure*}

\section{System under investigation}
The performance of the proposed NLC method was  evaluated through numerical simulations. All the investigated system scenarios are shown in Fig.~\ref{fig:setup}, with different colours representing different receiver schemes. 
An ideal optical transmitter was used to generate 5 polarization-multiplexed 16-ary quadrature amplitude modulated (PM-16QAM) channels at a symbol rate of 32 GBaud using a sequence of $2^{16}$ symbols. The signal pulses were shaped using a root raised cosine filter with 1\% roll-off factor. 
Transmission over $N_s$ 100~km-long optical fiber spans was simulated by numerically solving the Manakov equation using the split-step Fourier method. The fiber spans were based on SSMF and EDFA amplification after each span, with all parameters shown in Table~\ref{Tab. SimParameters} . At the mid-link point, the signal either continues to the following half of the link (blue path), or undergoes phase conjugation (red path). OPC was performed ideally by taking the complex conjugate of the optical signal in the time domain.        

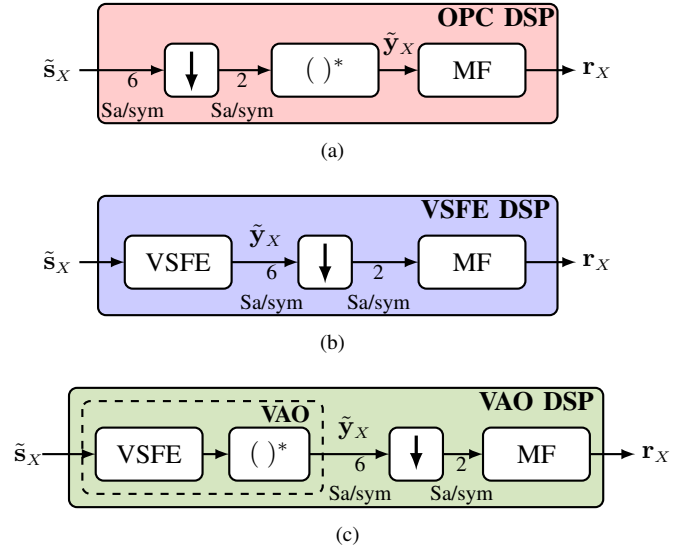
\begin{figure}
\centering
\subfloat[]{
\definecolor{mycolor5}{rgb}{0.46600,0.67400,0.18800}%
\begin{tikzpicture}
[plain/.style={align=center,execute at begin node=\setlength{\baselineskip}{2.5ex}}]
\draw[thick,rounded corners=3pt,fill=red,fill opacity=0.2] (115pt,20pt) rectangle (287pt,65pt);
\node[plain] at (100pt,40pt) {$\tilde{\mathbf{s}}_{\scaleto{X}{4pt}}$};
\draw[thick,->] (108pt,40pt) -- (140pt,40pt);
\draw[thick,rounded corners=3pt,fill=white] (140pt,30pt) rectangle (160pt,50pt);
\node[plain,align = center] at (128pt,30pt) {\footnotesize 6 \\ \footnotesize Sa/sym};
\node[plain,align = center] at (169pt,30pt) {\footnotesize 2 \\ \footnotesize Sa/sym};
\draw[very thick,->] (150pt,47pt) -- (150pt,32pt);
\draw[thick,rounded corners=3pt,fill=white] (180pt,30pt) rectangle (220pt,50pt);
\node[plain] at (200pt,40pt) {$(\;)^*$};
\draw[thick,->] (160pt,40pt) -- (180pt,40pt);
\draw[thick,->] (220pt,40pt) -- (235pt,40pt);
\draw[thick,rounded corners=3pt,fill=white] (235pt,30pt) rectangle (275pt,50pt);
\node[plain] at (255pt,40pt) {MF};
\draw[thick,->] (275pt,40pt) -- (294pt,40pt);
\node[plain] at (264pt,60pt) {\textbf{OPC DSP}};
\node[plain] at (228pt,50pt) {$\tilde{\mathbf{y}}_{\scaleto{X}{4pt}}$};
\node[plain] at (302pt,40pt) 
{$\mathbf{r}_{\scaleto{X}{4pt}}$};
\end{tikzpicture}
\label{subfig:OPC_DSP}
} \\
\subfloat[]{
\definecolor{mycolor5}{rgb}{0.46600,0.67400,0.18800}%
\begin{tikzpicture}
[plain/.style={align=center,execute at begin node=\setlength{\baselineskip}{2.5ex}}]
\draw[thick,fill=none,rounded corners=3pt,fill=blue,fill opacity=0.2] (115pt,-30pt) rectangle (287pt,15pt);
\node[plain] at (100pt,-10pt) {$\tilde{\mathbf{s}}_{\scaleto{X}{4pt}}$};
\draw[thick,->] (108pt,-10pt) -- (125pt,-10pt);
\draw[thick,rounded corners=3pt,fill=white] (125pt,-20pt) rectangle (165pt,0pt);
\node[plain] at (145pt,-10pt) {VSFE};
\draw[thick,->] (165pt,-10pt) -- (190pt,-10pt);
\draw[thick,rounded corners=3pt,fill=white] (190pt,-20pt) rectangle (210pt,0pt);
\draw[very thick,->] (200pt,-3pt) -- (200pt,-18pt);
\node[plain,align = center] at (180pt,-20pt) {\footnotesize 6 \\ \footnotesize Sa/sym};
\node[plain,align = center] at (220pt,-20pt) {\footnotesize 2 \\ \footnotesize Sa/sym};
\draw[thick,->] (210pt,-10pt) -- (235pt,-10pt);
\draw[thick,rounded corners=3pt,fill=white] (235pt,-20pt) rectangle (275pt,0pt);
\node[plain] at (260pt,10pt) {\textbf{VSFE DSP}};
\node[plain] at (255pt,-10pt) {MF};
\draw[thick,->] (275pt,-10pt) -- (294pt,-10pt);
\node[plain] at (178pt,0pt) {$\tilde{\mathbf{y}}_{\scaleto{X}{4pt}}$};
\node[plain] at (302pt,-10pt) {$\mathbf{r}_{\scaleto{X}{4pt}}$};
\end{tikzpicture}
\label{subfig:VSFE_DSP}
}\\
\subfloat[]{
\definecolor{mycolor5}{rgb}{0.46600,0.67400,0.18800}%
\begin{tikzpicture}
[plain/.style={align=center,execute at begin node=\setlength{\baselineskip}{2.5ex}}]
\draw[thick,rounded corners=3pt,fill=mycolor5,fill opacity=0.3] (100pt,-35pt) rectangle (300pt,-80pt);
\draw[thick,rounded corners=3pt,fill=none, dashed] (105pt,-40pt) rectangle (195pt,-75pt);
\node[plain] at (85pt,-60pt) {$\tilde{\mathbf{s}}_{\scaleto{X}{4pt}}$};
\draw[thick,->] (90pt,-60pt) -- (110pt,-60pt);
\draw[thick,rounded corners=3pt,fill=white] (110pt,-70pt) rectangle (150pt,-50pt);
\node[plain] at (130pt,-60pt) {VSFE};
\draw[thick,->] (150pt,-60pt) -- (160pt,-60pt);
\draw[thick,rounded corners=3pt,fill=white] (160pt,-70pt) rectangle (190pt,-50pt);
\node[plain] at (175pt,-60pt) {$(\;)^*$};
\node[plain] at (181pt,-45pt) {\small \textbf{VAO}};
\draw[thick,->] (190pt,-60pt) -- (220pt,-60pt);
\draw[thick,rounded corners=3pt,fill=white] (220pt,-70pt) rectangle (240pt,-50pt);
\draw[very thick,->] (230pt,-53pt) -- (230pt,-66pt);
\draw[thick,rounded corners=3pt,fill=white] (255pt,-70pt) rectangle (295pt,-50pt);
\node[plain] at (275pt,-60pt) {MF};
\node[plain] at (275pt,-40pt) {\textbf{VAO DSP}};
\draw[thick,->] (295pt,-60pt) -- (312pt,-60pt);
\draw[thick,->] (240pt,-60pt) -- (255pt,-60pt);
\node[plain,align = center] at (209pt,-70pt) {\footnotesize 6 \\ \footnotesize Sa/sym};
\node[plain,align = center] at (247pt,-70pt) {\footnotesize 2 \\ \footnotesize Sa/sym};
\node[plain] at (207pt,-50pt) {$\tilde{\mathbf{y}}_{\scaleto{X}{4pt}}$};
\node[plain] at (320pt,-60pt) 
{$\mathbf{r}_{\scaleto{X}{4pt}}$};
\end{tikzpicture}
\label{subfig:VAO_DSP}
}
\caption{Schematic diagram of the DSP chains for: \protect\subref{subfig:OPC_DSP} OPC-only system, \protect\subref{subfig:VSFE_DSP} conventional VSFE and \protect\subref{subfig:VAO_DSP} VAO. MF denotes the matched filtering stage and ${\mathbf{r}}_{\scaleto{X}{4pt}}$ the output symbols at 1 sample/symbol.}
\label{fig:NLC_DSP}
\end{figure}

After transmission, an optical coherent receiver was assumed to ideally capture the entire 5-channel optical bandwidth ($\approx$ 165 GHz) with a  subsequent sampling at 1.5 times the Nyquist rate of the received signal. The sampled signal over each polarization had 5 possible DSP chain options depending on the selected system (with or without OPC), which are: (i) ideal electronic dispersion compensation (EDC) performed in the frequency-domain when no OPC module was used; (ii) OPC-only DSP ; (iii) single-step/recursive VSFE; (iv) the VAO equalizer; (v) ideal NLC performed using a full-complexity (100 steps/span), \textit{full-field} DBP implementation. The DSP chain for each of the 3 NLC schemes compared in this work, i.e. OPC-only DSP, conventional VSFE, and VAO, is shown in Figs.~\ref{subfig:OPC_DSP}--\ref{subfig:VAO_DSP}. For each DSP scheme shown in Figs.~\ref{subfig:OPC_DSP}--\ref{subfig:VAO_DSP}, the input vector (shown only for the $X$ polarization for the sake of simplicity) is sampled at 6 samples/symbol. The OPC DSP module consists of a downsampling stage at 2 samples/symbol to select the central channel bandwidth, followed by time-domain conjugation, matched filtering and additional downsampling at 1 sample/symbol. For the conventional VSFE DSP, shown in Fig. \ref{subfig:VSFE}, the signal is first processed by the Volterra equalizer, downsampled at 2 samples/symbol and passed through a MF followed by downsampling to 1 sample/symbol. Similar operation is performed in the VAO DSP chain where, however, the signal must be conjugated after the VSFE block. After each DSP scheme, the SNR was estimated using a fully data-aided method. 

\begin{table}[b]
\centering
\normalsize
\caption{Simulation parameters}
\begin{tabular}{l|l}
\hline
Parameter     								   & Value \\ \hline
Symbol rate 								   & 32 GBd \\ \hline
Channel spacing								   & 32.5 GHz \\ \hline	
RRC roll-off	 	                           & 1 \% \\ \hline
Span length ($L_s$)							   & 100 km \\ \hline
Attenuation coefficient 					   & 0.20 dB \\ \hline
Dispersion parameter (D)               		   & 17 $\textnormal{ps}\cdot \textnormal{nm}^{-1}\cdot \textnormal{km}^{-1}$ \\ \hline
Nonlinear coefficient ($\gamma$)       		   & 1.2 W$^{-1}$km$^{-1}$ \\ \hline
EDFA noise figure							   & 5 dB  \\ \hline
\end{tabular}
\label{Tab. SimParameters}
\end{table}

\section{Nonlinearity suppression}
\label{sec:nonlinearity_suppression}
In order to assess the NLC effectiveness of the proposed scheme, a nonlinear interference (NLI) suppression factor was characterised through numerical simulations. This NLC suppression factor was defined as 
\begin{equation}
\zeta\triangleq\frac{\text{SNR}_{\text{\tiny{NLC}}}}{\text{SNR}_{\text{\tiny{EDC}}}},
\end{equation}
where $\text{SNR}_{\text{\tiny{EDC}}}$ and $\text{SNR}_{\text{\tiny{NLC}}}$ represent the SNR when EDC or NLC are applied, respectively, in the absence of ASE noise in the simulated system. The $\zeta$ indicates the reduction of the NLI power after NLC, and is key to understand, in the absence of any other noise sources, what is the effectiveness of that particular scheme. For an ideal NLC scheme (full compensation of any deterministic nonlinear effect), $\zeta$ tends to infinity as the NLI power is forced to 0. Conversely, this value tends to 1 if the NLI is left uncompensated (EDC-only schemes), or it is less than 1 when additional distortions are added, due for instance to NLC schemes operating with a very limited accuracy. In order to characterize the proposed VAO scheme, the $\zeta$ parameter was studied as a function of the 2 optical system parameters: transmitted power, and transmission distance for a 5-channel transmission. $\zeta$ was calculated numerically through Monte-Carlo estimation of the received SNR, in the absence of ASE noise in the simulated system. An appropriate number of processed subsequences was accumulated to guarantee an SNR accuracy within 0.05 dB.       

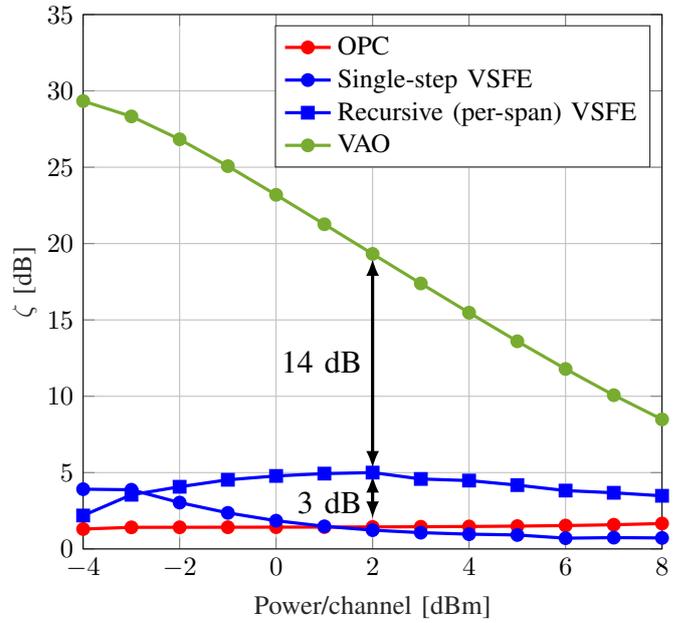
\begin{figure}[!t]
\definecolor{mycolor5}{rgb}{0.46600,0.67400,0.18800}%

\begin{tikzpicture}
\begin{axis}[
width=0.86\columnwidth,
height=0.8\columnwidth,
at={(0.769in,0.47in)},
scale only axis,
xmin=-4,
xmax=8,
xlabel style={font=\color{white!15!black}},
xlabel={Power/channel [dBm]},
ymin=0,
ymax=35,
ylabel style={font=\color{white!15!black}},
ylabel={$\zeta$ [dB]},
ylabel shift =-3pt,
axis background/.style={fill=white},
xmajorgrids,
ymajorgrids,
legend style={legend cell align=left, align=left, draw=white!15!black}
]

\addplot [color=red,very thick,mark=*] file{Figures/Data/NLsupp_vs_Pow_OPC_5ch_1000km.txt};\addlegendentry{OPC};
    
\addplot [color=blue,very thick,mark=*]  file{Figures/Data/NLsupp_vs_Pow_Volterra_5ch_1000km.txt};\addlegendentry{Single-step VSFE};
 
\addplot [color=blue,very thick,mark=square*]  file{Figures/Data/NLsupp_vs_Pow_Enh_Recursive_Volterra_5ch_1000km.txt
};\addlegendentry{Recursive (per-span) VSFE};

\addplot [color=mycolor5,very thick,mark=*]  file{Figures/Data/NLsupp_vs_Pow_OPC+Volterra_5ch_1000km_2.txt};\addlegendentry{VAO};

\draw[<->,very thick] (axis cs:2,5.3)--++(0pt,40pt) node[left] {\large 14 dB}--(axis cs:2,19);
\draw[<->,very thick] (axis cs:2,4.8)--++(0pt,-10pt) node[left] {\large 3 dB}--(axis cs:2,2);

\end{axis}
\end{tikzpicture}
\vspace{-0.4cm}
\caption{Nonlinearity suppression factor ($\zeta$) as a function of transmitted power for the NLC schemes studied in this work.}
\label{fig:NLsupp_vs_pow}
\end{figure}

The nonlinearity suppression factor ($\zeta$) was calculated as a function of power for the transmission of a 5$\times$32 GBaud signal over a 1000 km link for 3 different NLC schemes illustrated in Fig.~\ref{fig:NLC_DSP}: mid-link OPC (red curve); VSFE (blue curves) and the proposed VAO (green curve). These results are plotted in Fig.~\ref{fig:NLsupp_vs_pow}. The VSFE is here implemented in two different variants: single-step (blue curve with circle markers), and recursive VSFE (squared markers). 
Both variants of VSFE and VAO were operated (repeatedly) over 512 symbol-long sequences each oversampled at 6 samples/symbol, resulting in 3072 samples per processed sequence.   
The results show that OPC (red curve) has a constant NL suppression of approximately 1.8 dB across the entire range of powers shown. Such a limited NLC suppression is the result of the residual NLI when OPC is used in an EDFA link (see Sec.~\ref{sec:VAO}), and is explained by the third-order Volterra kernel illustrated in Fig.~\ref{subfig:kernel_b}. 
VSFE was characterized in terms of $\zeta$ in the two different implementations described in Sec.~\ref{sec:VAOimplementation}: single-step and recursive VSFE. Single-step VSFE shows a better performance at low powers, with a maximum $\zeta$=2 dB at -4 dBm, and monotonically decreasing $\zeta$ as the transmitted power increases. This well-known performance behaviour of the Volterra equalizer is due to the truncation of the Volterra series to the third-order term. Such a truncation becomes less accurate as the transmitted power is increased, and leads to a degradation in the equalization performance ($\zeta$=0.5 dB at 8 dBm/channel). This degradation is partly mitigated when a recursive VSFE scheme is used. Although the two VSFE schemes (blue curves in Fig.~\ref{fig:NLsupp_vs_pow}) show similar NLI suppression ($\approx$ 3.5 dB) for around -3 dBm/channel, the recursive VSFE scheme performs better beyond -3 dBm/channel and achieves up to 5 dB NLI suppression at 2 dBm/channel. However, this improvement comes at the expense of an $N_s$ times higher computational complexity compared to a single-step VSFE scheme, for a fixed processed sequence length \cite{Guiomar2013}. Fig.~\ref{fig:NLsupp_vs_pow} shows that the VAO scheme (green curve) has a superior performance compared to the all the other NLC schemes analysed, across the entire range of powers. Alike single-step VSFE, the VAO $\zeta$ factor monotonically decreases with power, and reaches a maximum value (within the range of powers shown in Fig.~\ref{fig:NLsupp_vs_pow}) of $\approx$ 30~dB at -5~dBm/channel. At this transmitted power, VAO exceeded the NLI suppression of all of the other NLC schemes by over 25~dB. Although the performance of VAO can be observed to degrade quickly as a function of power, even in the region of powers of interest for optimum transmission performance (0 to 4 dBm) the VAO NLI suppression is still significantly higher than other compared schemes. 
In particular, at 2 dBm/channel VAO achieves 17 dB higher $\zeta$ compared to using only OPC, and 14 dB higher suppression than the optimum $\zeta$ achieved by a VSFE-only scheme ($\zeta$=5 dB at 2 dBm/channel).    

The remarkable NLC performance attained by the VAO for the system studied is attributed to two reasons: i) for a fixed processed sequence length, VAO benefits from a reduced channel memory compared to VSFE; ii) the adoption of OPC leads to a more accurate reconstruction of the transmitted field using a third-order Volterra series term due to the reduced NLI at the receiver. The reduction of the channel memory is due to the compensating effect of the phase conjugation in the second part of the link as the OPC completely removes the fibre chromatic dispersion accumulated in the first part of the link, thus also reducing the time window over which the nonlinear intersymbol takes place. This \emph{memory reduction} can also be inferred by the phased-array in \eqref{eq:VolterraInt_OPC} which is only evaluated over half of the link. 

\begin{figure*}[!t]
\subfloat[]{
\definecolor{mycolor5}{rgb}{0.46600,0.67400,0.18800}%

\begin{tikzpicture}
\hspace{-0.4cm}
\begin{axis}[
width=0.6\columnwidth,
height=0.55\columnwidth,
at={(0.769in,0.47in)},
scale only axis,
xmin=0,
xmax=5000,
xtick={0,1000,2000,3000,4000,5000},
xticklabels={0,1000,2000,3000,4000,5000},
xlabel style={font=\color{white!15!black}},
xlabel={Transmission Distance [km]},
ymin=0,
ylabel style={font=\color{white!15!black}},
ylabel={$\zeta$ [dB]},
ylabel shift =-7pt,
axis background/.style={fill=white},
xmajorgrids,
ymajorgrids,
legend style={legend cell align=left, align=left, draw=white!15!black}
]

\addplot [color=purple,very thick,mark=square*]  file{Figures/Data/NLsupp_vs_Dist_Volterra_5ch_0dBm_128sym.txt};\addlegendentry{128 sym};

\addplot [color=mycolor5,very thick,mark=square*]  file{Figures/Data/NLsupp_vs_Dist_Volterra_5ch_0dBm_256sym.txt};\addlegendentry{256 sym};

\addplot [color=red,very thick,mark=square*]  file{Figures/Data/NLsupp_vs_Dist_Volterra_5ch_0dBm_512sym.txt};\addlegendentry{512 sym};

\addplot [color=blue,very thick,mark=square*]  file{Figures/Data/NLsupp_vs_Dist_Volterra_5ch_0dBm_1024sym.txt};\addlegendentry{1024 sym};

\end{axis}
\end{tikzpicture}
\label{subfig:NLsupp_vs_Dist_a}}
\subfloat[]{\hspace{-1.1cm}
\definecolor{mycolor5}{rgb}{0.46600,0.67400,0.18800}%

\begin{tikzpicture}
\begin{axis}[
width=0.6\columnwidth,
height=0.55\columnwidth,
at={(0.769in,0.47in)},
scale only axis,
xmin=0,
xmax=5000,
xlabel style={font=\color{white!15!black}},
xlabel={Transmission Distance [km]},
xtick={0,1000,2000,3000,4000,5000},
xticklabels={0,1000,2000,3000,4000,5000},
ymin=0,
ylabel style={font=\color{white!15!black}},
ylabel shift =-3pt,
axis background/.style={fill=white},
xmajorgrids,
ymajorgrids,
legend style={legend cell align=left, align=left, draw=white!15!black}
]

\addplot [color=purple,very thick,mark=*]  file{Figures/Data/NLsupp_vs_Dist_VOA_5ch_0dBm_128sym.txt};\addlegendentry{128  sym};

\addplot [color=mycolor5,very thick,mark=*]  file{Figures/Data/NLsupp_vs_Dist_VOA_5ch_0dBm_256sym.txt};\addlegendentry{256 sym};

\addplot [color=red,very thick,mark=*]  file{Figures/Data/NLsupp_vs_Dist_VOA_5ch_0dBm_512sym.txt};\addlegendentry{512 sym};

\addplot [color=blue,very thick,mark=*]  file{Figures/Data/NLsupp_vs_Dist_VOA_5ch_0dBm_1024sym.txt};\addlegendentry{1024 sym};
  
\draw[<->,very thick] (axis cs:2000,11.7)--(axis cs:2000,19.4); 
\node[fill=white] (label1) at (axis cs:2750,14) {\large 8.6 dB};
\draw[<->,very thick] (axis cs:4900,2.5)--(axis cs:4900,7.2); 
\node[fill=white] (label2) at (axis cs:4300,6) {\large 6 dB};

\end{axis}
\end{tikzpicture}
\label{subfig:NLsupp_vs_Dist_b}}
\subfloat[]{\hspace{-0.8cm}
\definecolor{mycolor5}{rgb}{0.46600,0.67400,0.18800}%

\begin{tikzpicture}
\begin{axis}[
width=0.6\columnwidth,
height=0.55\columnwidth,
at={(0.769in,0.47in)},
scale only axis,
xmin=0,
xmax=5000,
xlabel style={font=\color{white!15!black}},
xlabel={Transmission Distance [km]},
xtick={0,1000,2000,3000,4000,5000},
xticklabels={0,1000,2000,3000,4000,5000},
ymin=0,
ymax=30,
ylabel style={font=\color{white!15!black}},
ylabel={$\zeta$ gain [dB]},
ylabel shift =-6pt,
axis background/.style={fill=white},
xmajorgrids,
ymajorgrids,
legend style={legend cell align=left, align=left, draw=white!15!black,at={(0.47,0.4)}}
]

\addplot [color=purple,very thick,mark=*]  file{Figures/Data/VAO_NLsupp_gain_vs_Dist_128sym.txt};\addlegendentry{128 sym};

\addplot [color=mycolor5,very thick,mark=*]  file{Figures/Data/VAO_NLsupp_gain_vs_Dist_256sym.txt};\addlegendentry{256 sym};

\addplot [color=red,very thick,mark=*]  file{Figures/Data/VAO_NLsupp_gain_vs_Dist_512sym.txt};\addlegendentry{512 sym};

\addplot [color=blue,very thick,mark=*]  file{Figures/Data/VAO_NLsupp_gain_vs_Dist_1024sym.txt};\addlegendentry{1024 sym};

  \end{axis}
\end{tikzpicture}
\label{subfig:NLsupp_vs_Dist_c}}

\caption{Nonlinearity suppression factor ($\zeta$) as a function of transmission distance and different equalizer window sizes for \protect\subref{subfig:NLsupp_vs_Dist_a} VSFE, \protect\subref{subfig:NLsupp_vs_Dist_b} VAO, and \protect\subref{subfig:NLsupp_vs_Dist_c} VAO NLI suppression gain compared to VSFE.}
\label{fig:NLsupp_vs_Dist}
\end{figure*}
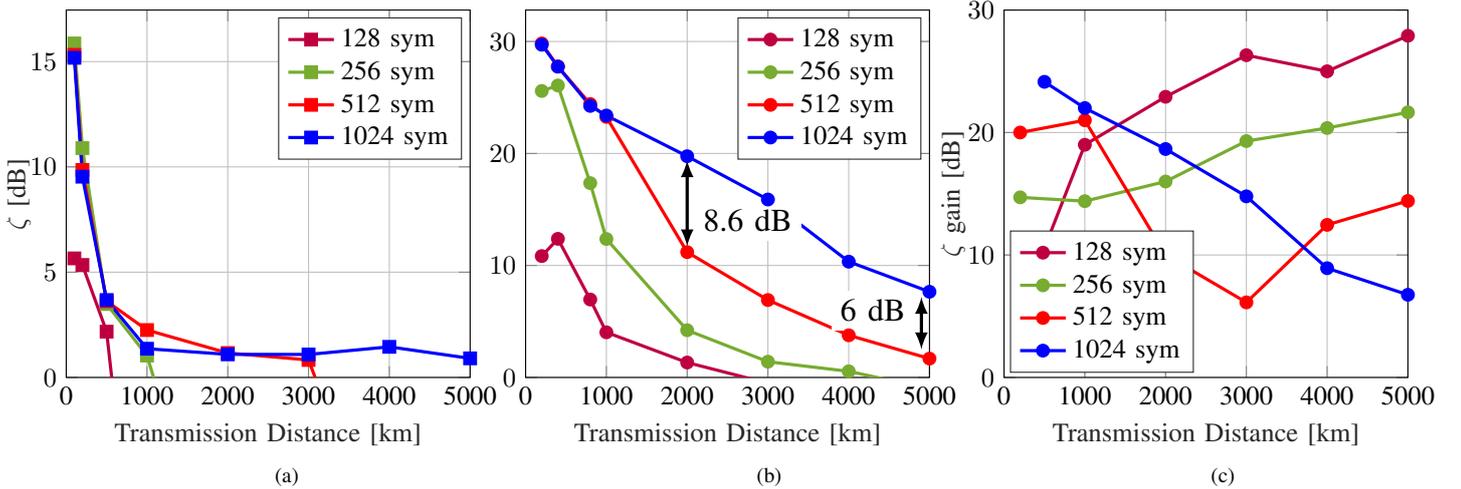

To assess the improved robustness of VAO against an increasing channel memory, the NLI suppression as a function of the transmitted distance is shown for the system under investigation at a transmitted power of 0~dBm/channel in Fig.~\ref{fig:NLsupp_vs_Dist}, for single-step VSFE (Fig.~\ref{subfig:NLsupp_vs_Dist_a}), and VAO (Fig.~\ref{subfig:NLsupp_vs_Dist_b}). The effect of the channel memory is highlighted by showing the performance for different window sizes, spanning from 128 to 1024  symbols. As the channel memory is expected to scale linearly with increasing transmission distance, each scheme NLC shows a decreasing NLI suppression factor for a fixed window size. However, performance degradation as a function of distance is also caused by an increasing dominance of higher-order Volterra series terms. The two effects can be distinguished by observing the behaviour of $\zeta$ as the window size is increased. The results in Fig.~\ref{subfig:NLsupp_vs_Dist_a} show that, up to 3000 km transmission distance, $\zeta$ saturates using time-windows between 512 and 1024 symbols. For each window size, a \emph{threshold distance} of operation, beyond which VSFE performance breaks down (negative $\zeta$), can be identified. This was caused by the channel memory going beyond the maximum number of discarded symbols within a given window size. Increasing from 512 to 1024 symbols the window size did not lead to any substantial increase in $\zeta$. Thus, it can be concluded that for window sizes beyond 1024 symbols and up to 5000 km transmission distance, the VSFE performance is dominated by its compensation accuracy (NLI compensation up to the first-order).   

The VAO NLI suppression as a function of distance is shown in Fig.~\ref{subfig:NLsupp_vs_Dist_b}. It can be observed that for VAO the detrimental effect of the channel memory operating over finite window sizes is less pronounced. This is also confirmed by the smoother decay of $\zeta$ as a function of transmission distance compared to VSFE, with NLC capabilities (positive $\zeta$) observed within 3000 km for all window sizes shown. For transmission distances up to 1000 km, the value of $\zeta$ saturates already at 512 symbols. Beyond a distance of 1000 km, the increase of the window size from 512 symbols to 1024 symbols results in a substantial increase in $\zeta$ (up to 8.6 dB at 2000 km). The performance gap between using a 512-symbol window and a 1024-symbol window then gradually decreases to 6 dB at 5000 km. This can be attributed to the stronger effect of higher-order Volterra series terms as the transmission distance increases.        
Fig.~\ref{subfig:NLsupp_vs_Dist_c} shows the NLI suppression gain achieved by VAO over VSFE as a function of distance for different window lengths. The power per channel is again fixed at 0 dBm/channel. It can be seen that the improvement introduced by the VAO scheme in terms of $\zeta$ is remarkable for all distances and window sizes used. For short distances, (up to 1000 km) the highest gains can be observed for 512- and 1024-symbol window length, up to 21 dB and 24 dB, respectively. This highlights the superior NLC efficacy of a Volterra-based equalizer when an in-line OPC device is used. For longer distances (beyond 2000 km), the highest VAO gain is achieved for the shortest window lengths (128- and 256-symbol long), due to the reduction of the channel memory provided by OPC. In contrast with VSFE, this enables the VAO equalizer to obtain a certain (albeit limited) amount of NLI suppression ($\leq 4.5$ dB) for distances within $2000-4000$ km, while still processing relatively short sequences (see Fig.~\ref{subfig:NLsupp_vs_Dist_b}). For instance, a 19.5 dB gain is attained using VAO with a 256-symbol long sequence at 3000 km, which brings the VAO $\zeta$ factor up to $\approx$ 1.5 dB from a -18.5 dB achieved by the VSFE over the same window length (see Fig.~\ref{subfig:NLsupp_vs_Dist_a} and \ref{subfig:NLsupp_vs_Dist_b}).            

\section{SNR performance}
\label{sec:snr_performance}
%
In the previous section, the performance of the proposed VAO scheme was analysed in terms of NLI suppression factor. Despite this being a more direct measure of the NLC effectiveness, it does not immediately translate into more insightful performance metrics such as signal-to-noise ratio (SNR). In this section, a study of the SNR performance among the different NLC schemes studied in this work is described. 

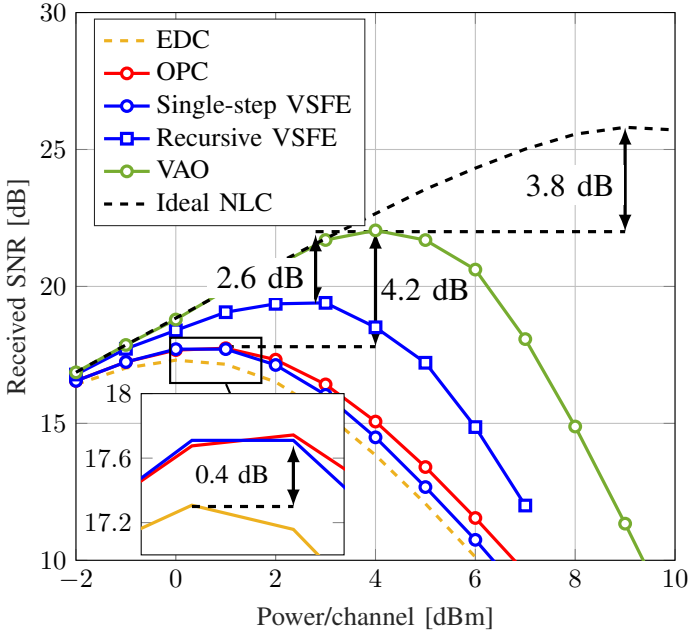
\begin{figure}[!t]
%
%
\definecolor{mycolor1}{rgb}{0.00000,0.44700,0.74100}%
\definecolor{mycolor2}{rgb}{0.85000,0.32500,0.09800}%
\definecolor{mycolor3}{rgb}{0.92900,0.69400,0.12500}%
\definecolor{mycolor4}{rgb}{0.49400,0.18400,0.55600}%
\definecolor{mycolor5}{rgb}{0.46600,0.67400,0.18800}%
\hspace{-0.3cm}
\begin{tikzpicture}
\begin{axis}[
width=0.89\columnwidth,
height=0.82\columnwidth,
at={(0.769in,0.47in)},
scale only axis,
xmin=-2,
xmax=10,
xlabel style={font=\color{white!15!black}},
xlabel={Power/channel [dBm]},
ymin=10,
ymax=30,
ylabel style={font=\color{white!15!black}},
ylabel={Received SNR [dB]},
ylabel shift=-3pt,
axis background/.style={fill=white},
xmajorgrids,
ymajorgrids,
legend style={at={(0.03,0.61)}, anchor=south west, legend cell align=left, align=left, draw=white!15!black},
clip mode=individual
]

\draw[<->,very thick] (axis cs:4,17.8)--(axis cs:4,22);
\draw[dashed,very thick] (axis cs:1,17.8)--(axis cs:4,17.8);
\node[fill=none] at (axis cs:5,19.9) {\large 4.2 dB}; 

\draw[<->,very thick,fill=white] (axis cs:9,22)--(axis cs:9,25.8);
\draw[dashed,very thick] (axis cs:2.8,22)--(axis cs:9,22);
\node[fill=white] at (axis cs:7.9,23.7) {\large 3.8 dB}; 


\addplot [color=mycolor3,very thick, dashed]
  table[row sep=crcr]{%
-2	16.424350498216\\
-1	17.0201386454314\\
0	17.3079638864184\\
1	17.1581747666859\\
2	16.5104594883734\\
3	15.3644543256797\\
4	13.8395878742004\\
5	12.0674659463316\\
6	10.1183176537121\\
7	8.04980105654656\\
8	5.84582520092793\\
9	3.48231251748383\\
10	0.861114258241587\\
};
\addlegendentry{EDC}

\addplot [color=red,very thick,mark=*, mark options={fill=white}]
  table[row sep=crcr]{%
-2	16.5514591048299\\
-1	17.2410276044596\\
0	17.6758780475559\\
1	17.7436190008049\\
2	17.3257805725543\\
3	16.4226219222562\\
4	15.0660511700697\\
5	13.4055437577421\\
6	11.5453009452506\\
7	9.56438923496832\\
8	7.45844696763862\\
9	5.23400585784181\\
10	2.83431937247963\\
};
\addlegendentry{OPC}

\addplot [color=blue,very thick, mark=*, mark options={fill=white}]
  table[row sep=crcr]{%
-2	16.5514591048299\\
-1	17.2410276044596\\
0	17.7101082778063\\
1	17.7101082778063\\
2	17.1322842649176\\
3	16.0415971119223\\
4	14.4888054513863\\
5	12.6741706937716\\
6	10.7472935458019\\
7	8.65084845370653\\
8	6.44625633198677\\
9	4.09111476054413\\
10	1.45256642120736\\
};
\addlegendentry{Single-step VSFE}

\addplot [color=blue,very thick, mark=square*, mark options={fill=white}]
  table[row sep=crcr]{%
-2	16.7838\\
-1	17.7222\\
0	18.3959\\
1	19.059\\
2	19.3648\\
3   19.3999\\
4	18.508\\
5 	17.2129\\
6 	14.8643\\
7   12\\
};
\addlegendentry{Recursive VSFE}

\addplot [color=mycolor5,very thick,mark=*, mark options={fill=white}]
  table[row sep=crcr]{%
-2	16.8632932645276\\
-1	17.863076715826\\
0	18.7917295371448\\
1	19.8637720817771\\
2	20.7390461659056\\
3	21.6948485276793\\
4	22.0464376854419\\
5	21.6925484679542\\
6	20.6149341326072\\
7	18.0772284440824\\
8	14.8849696750958\\
9	11.3383498614333\\
10	7.64748813665142\\
};
\addlegendentry{VAO}

\addplot [color=black,very thick,dashed]
  table[row sep=crcr]{%
-2	16.8632932645276\\
-1	17.863076715826\\
0	18.8430491497794\\
1	19.8244956383953\\
2	20.8099154298978\\
3	21.7461207072213\\
4	22.658197746364\\
5	23.5466385336708\\
6	24.3181770384092\\
7	25.0137279832617\\
8	25.5597546071734\\
9	25.8098525526889\\
10	25.7147143610926\\
};
\addlegendentry{Ideal NLC}
\node (rect) at (axis cs: 0.8,17.3)[draw,thick,minimum width=1.2cm,minimum height=0.6cm] (box) {};
\draw[thick] (box)--(axis cs: 1.3,15.2);
\draw[thick, dashed] (axis cs: 1.3,15.2)--(axis cs: 1.3,15.2);
\draw[very thick, <->] (axis cs: 2.8,19.35)--++(0pt,10pt) node[left,fill=white]{\large 2.6 dB}--(axis cs: 2.8,21.9);
\end{axis}

\begin{axis}[xshift=80pt,
yshift=36pt,
width=0.48\columnwidth,
height=0.42\columnwidth,
xmin=-0.5,
xmax=1.5,
ymin=17,
ymax=18,
axis background/.style={fill=white},
xmajorticks=false,
ytick={17.2,17.6,18}]

\addplot [color=mycolor3,very thick]
  table[row sep=crcr]{%
-2	16.424350498216\\
-1	17.0201386454314\\
0	17.3079638864184\\
1	17.1581747666859\\
2	16.5104594883734\\
3	15.3644543256797\\
4	13.8395878742004\\
5	12.0674659463316\\
6	10.1183176537121\\
7	8.04980105654656\\
8	5.84582520092793\\
9	3.48231251748383\\
10	0.861114258241587\\
};

\addplot [color=red,very thick]
  table[row sep=crcr]{%
-2	16.5514591048299\\
-1	17.2410276044596\\
0	17.6758780475559\\
1	17.7436190008049\\
2	17.3257805725543\\
3	16.4226219222562\\
4	15.0660511700697\\
5	13.4055437577421\\
6	11.5453009452506\\
7	9.56438923496832\\
8	7.45844696763862\\
9	5.23400585784181\\
10	2.83431937247963\\
};

\addplot [color=blue,very thick]
  table[row sep=crcr]{%
-2	16.5514591048299\\
-1	17.2410276044596\\
0	17.7101082778063\\
1	17.7101082778063\\
2	17.1322842649176\\
3	16.0415971119223\\
4	14.4888054513863\\
5	12.6741706937716\\
6	10.7472935458019\\
7	8.65084845370653\\
8	6.44625633198677\\
9	4.09111476054413\\
10	1.45256642120736\\
};
\draw[<->,very thick] (axis cs:1,17.3)--(axis cs:1,17.68);
\draw[dashed,very thick] (axis cs:0,17.3)--(axis cs:1,17.3);
\node[fill=white] at (axis cs:0.4,17.5) {0.4 dB}; 
\end{axis}
\end{tikzpicture}%
\caption{Received SNR for different NLC schemes as a function of signal launch power.
\label{subfig:SNR_power}
}
\end{figure}

\begin{figure}[!t]
%
%
\definecolor{mycolor1}{rgb}{0.00000,0.44700,0.74100}%
\definecolor{mycolor2}{rgb}{0.85000,0.32500,0.09800}%
\definecolor{mycolor3}{rgb}{0.92900,0.69400,0.12500}%
\definecolor{mycolor4}{rgb}{0.49400,0.18400,0.55600}%
\definecolor{mycolor5}{rgb}{0.46600,0.67400,0.18800}%
\begin{tikzpicture}

\begin{axis}[%
width=0.89\columnwidth,
height=0.82\columnwidth,
at={(0.769in,0.47in)},
scale only axis,
xmin=0,
xmax=3500,
xlabel style={font=\color{white!15!black}},
xlabel={Distance [km]},
ymin=10,
ymax=35,
xtick={0,500,1000,1500,2000,2500,3000,3500},
xticklabels={0,500,1000,1500,2000,2500,3000,3500},
ylabel style={font=\color{white!15!black}},
ylabel={Received SNR [dB]},
ylabel shift=-3pt,
axis background/.style={fill=white},
xmajorgrids,
ymajorgrids,
legend style={at={(0.97,0.61)}, anchor=south east, legend cell align=left, 
align=left, draw=white!15!black}
]
\draw[<->,very thick] (axis cs:600,20)--(axis cs:1450,20);
\draw[<->,very thick] (axis cs:1530,20)--(axis cs:2450,20);
\addplot [color=mycolor3,very thick]
  table[row sep=crcr]{%
200 24.79\\
400	21.4978907240954\\
800	18.3382157494481\\
1200	16.482696670017\\
1600	15.1788978430153\\
2000	14.1559210132702\\
2600	12.9339155293656\\
3200	11.9976340954077\\
4000	10.9436311674154\\
5000	9.87360840024922\\
6000	8.98107231471431\\
};
\addlegendentry{EDC}

\addplot [color=red,very thick]
  table[row sep=crcr]{%
200 24.99\\
400	21.8170775625406\\
800	18.7496081641795\\
1200	16.9293654015706\\
1600	15.6478932517892\\
2000	14.684287830246\\
2600	13.4770765141391\\
3200	12.5632675443772\\
4000	11.5369422589118\\
5000	10.4952247073809\\
6000	9.67773526337443\\
};
\addlegendentry{OPC}

\addplot [color=blue,very thick]
  table[row sep=crcr]{%
200 26.9129\\  
400	22.6710201436397\\
800	18.9873017712903\\
1200	16.8786729375625\\
1600	15.4609\\
2000	14.1187\\
2600	12.9384091548841\\
3200	11.8618040779372\\
};
\addlegendentry{Single-step VSFE}

\addplot [color=blue,very thick,dashed]
  table[row sep=crcr]{%
200 27.6948\\  
400	24.6656\\
800	20.8752\\
1200	18.0205\\
1600	15.46566\\
};
\addlegendentry{Recursive VSFE}

\addplot [color=mycolor5,very thick]
  table[row sep=crcr]{%
200 30.7838\\
400	27.0229048140704\\
800	23.2391258547945\\
1200	21.2544661375941\\
1600	19.317826732644\\
2000	17.8384567841875\\
2600	15.9054242603843\\
3200	14.5497368932477\\
4000	13.0136511596142\\
5000	11.5458635140069\\
6000	10.4868790790133\\
};
\addlegendentry{VAO}

\addplot [color=black,very thick,dashed]
  table[row sep=crcr]{%
200 34.33  
400	30.9406794143077\\
800	27.1901738265184\\
1200	24.6733952284495\\
1600	22.7823681459583\\
2000	21.3153510869646\\
2600	19.6296234893415\\
3200	18.259038277325\\
4000	16.7107182894681\\
5000	15.2652841361631\\
6000	14.0354785919831\\
};
\addlegendentry{Ideal NLC}
\node[fill=white] at (axis cs:1100,21.5) {\large +200 $\%$}; 
\node[fill=white] at (axis cs:2000,21.5) {\large +66 $\%$}; 
\end{axis}
\end{tikzpicture}%
\caption{Received SNR for different NLC schemes as a function of transmission distance.}
\label{subfig:SNR_Distance}
\end{figure}
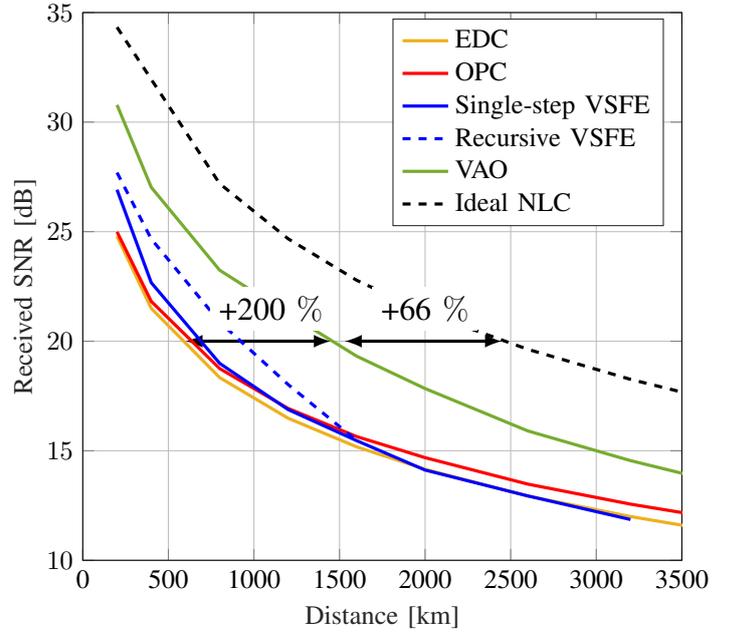

Numerical simulations were carried out to assess the performance of different NLC methods in a realistic transmission scenario and to compare their performance with the VAO method. Similar to Sec.~\ref{sec:nonlinearity_suppression}, the performance was first evaluated for a fixed transmission distance of 1000~km as a function of the signal launch power, with results shown in Fig.~\ref{subfig:SNR_power}, and subsequently as a function of the transmission distance at optimum power up to 3200~km, presented in Fig.~\ref{subfig:SNR_Distance}.

Figure~\ref{subfig:SNR_power} shows the performance of the system using the different NLC techniques as a function of launch power per channel after transmission of 1000~km. The use of electronic dispersion compensation (EDC) shows the worst performance among all the studied methods. As EDC does not enact any compensation of nonlinear effects, this scheme serves as a baseline to evaluate the effectiveness of the NLC methods under study. EDC exhibits a maximum SNR of 17.3~dB at 0~dBm per channel. At this distance, OPC presents a gain over EDC limited to 0.4~dB, with a maximum SNR of 17.7~dB, indicating that only a small portion of the nonlinearities are compensated. As discussed in Sec.~ \ref{sec:VAO}, the spectral region where OPC offers complete cancellation of nonlinearites (see Fig.~\ref{subfig:kernel_b}) was not sufficiently large to account for the effects of all 5 transmitted channels. In the highly-nonlinear regime, e.g. using a power of 4~dBm, an increase of 1.2~dB in the SNR over EDC is observed with the use of OPC. This value is in agreement with the nonlinear suppression factor observer for OPC in the previous section. Similar performance is observed using single-step VSFE at the maximum SNR. In this case, one-step VSFE was implemented using a window size of 512 symbols. A maximum SNR of 17.7~dB is found using this method at a signal power of 1~dBm. 
As the power is increased, the efficiency of VSFE is decreased as a result of the truncation of the Volterra series to third-order term, as described in the previous section. 
For the recursive VSFE a window size of 256 symbols was used. This method presents an improved performance relative to the one-step implementation, with a maximum SNR of 19.4~dB at 2~dBm/channel signal power. 
Using VAO results in a significant improvement relative to VSFE and OPC, with a gain of 4.6~dB compared to EDC, 4.2~dB relative to either conventional one-step VSFE or OPC, and  2.6~dB compared to recursive VSFE. The VAO was implemented using a window size of 512 symbols in the equalizer, to allow a direct comparison with the conventional VSFE methods. Whilst the VSFE truncation is one of the factors that limited the performance of conventional VSFE implementations, the reduction of nonlinear effects due to the mid-link OPC leads to a better reconstruction of the transmitted signal using the VAO scheme.
Finally, for comparison, the performance of ideal NLC is shown, obtained by perfectly integrating the Manakov equation in a virtually reversed transmission link. This represents the maximum performance that can be obtained using any receiver based NLC scheme. Although the VAO scheme presents improved performance relative the aforementioned NLC schemes, a 3.8~dB gap is observed compared to an ideal NLC method.

Figure~\ref{subfig:SNR_Distance} shows the performance of the system as a function of transmission distance. The same trend from Fig.~\ref{subfig:SNR_power} in terms of nonlinear effects was observed, with EDC exhibiting the worst performance among all the studied methods. 
Interestingly, OPC presents a reduced gain of 0.2~dB relative to EDC after 200~km, however greater nonlinearity compensation is obtained for longer distances and 0.6~dB gain is observed after 3200~km.
The opposite behaviour is found for single-step VSFE, where a superior performance is found at short distances. As in the previous case, both standard VSFE were implemented using window sizes of 512 and 256 symbols for the single-step and recursive algorithms, respectively. 
Using VSFE, the gain over EDC is slowly reduced as the transmission distance increased, up to the point where no improvement is observed. This phenomenon can be understood from memory effects introduced by the fiber chromatic dispersion. When the temporal spreading of the propagated signal exceeds the VSFE window size the performance of both implementations, single-step and recursive, is dramatically reduced. For the studied system no gain from NLC is observed for transmission distances longer than 1600~km using both standard VSFE implementations. 

The VAO scheme is shown to outperform both standard implementations of VSFE, and OPC at all evaluated transmission distances. For a received SNR of 20~dB, VOA is shown to attain a 200\% increase in transmission distance compared to EDC, going from 500 to 1500~km. The VAO method however, did not offer full compensation of the NLI experienced during transmission and reduced performance is found compared to ideal NLC. The use of ideal NLC achieves an extra 66\% increase of transmission distance for a received SNR of 20~dB compared to VAO. 
Additionally, the NLC gain from the VAO method presents an accelerated reduction as a function of the transmission distance, compared to ideal NLC. We attribute this behaviour to the nonlinear equalizer in the VAO scheme, which has qualitatively the same limitations as a standard VSFE in terms of channel memory and truncation of the Volterra series. 

\section{Complexity}
\label{sec:complexity}
In the previous sections, the VAO scheme was shown to significantly outperform OPC and VSFE in practical transmission scenarios, such as EDFA-amplified links, for a NLC bandwidth of 165 GHz (5$\times$32 GBaud). Although VAO offers an improved performance than VSFE, a substantial performance penalty (3.8 dB at 1000 km transmission distance) was still observed with respect to ideal nonlinearity compensation (see Fig.~\ref{subfig:SNR_power}).
Such an ideal NLC performance could be theoretically achieved through a \emph{full-complexity} implementation of the DBP algorithm over the entire transmitted bandwidth. The question that naturally arises is then, which, between VAO and DBP, represents a better trade-off between performance and complexity.

The computational complexity of the VAO scheme proposed in this work is determined by the complexity of its modified Volterra equalizer. Comprehensive studies on the complexity of VSFE, and simplified variants thereof, can be found in literature \cite{Weidenfeld2010, Guiomar2011, Guiomar2013, Liu2012, Shulkind2013, Bakhshali2016}. These works showed that, despite the complexity (per processed symbol) of a standard VSFE implementations scaling asymptotically as $\mathcal{O}(N^2)$, simplified implementations are possible with limited performance penalties. In particular, in \cite{Guiomar2013} and \cite{Shulkind2013} the complexity of VSFE was reduced to $\mathcal{O}(N)$ and $\mathcal{O}(N\log N)$ per symbol, respectively.

The computational complexity of the DBP algorithm, for a large number of processed samples $N$, scales as $\mathcal{O}(N_{\text{steps}}\log N)$ where $N_{\text{steps}}$ is the number of DBP steps performed. 
Alike the VSFE/VAO case discussed in Sec.~\ref{sec:nonlinearity_suppression}, the minimum value of $N$ for DBP is dictated by the targeted NLC performance and the channel memory (see e.g.~\cite{Secondini2016}). Due the logarithmic scaling of simplified VSFE, matching an increasing channel memory by increasing $N$ only results in a minor complexity increase, 
which is also the case for DBP. 
The main issue in extending DBP to larger NLC bandwidths is the minimum $N_{\text{steps}}$ required to preserve a fixed NLC suppression. The NLC performance for DBP is determined by the relationship between the accuracy of the SSFM and the signal bandwidth at a fixed power spectral density, which depends on the specific SSFM method used and, overall, it is not well known. For instance, for a logarithmic step-size SSFM \cite{Bosco2000}, $N_{\text{steps}}$ scales quadratically with the NLC bandwidth (at fixed power spectral density). Although for other SSFM variants (see, e.g., \cite{Secondini2016}) this scaling could be more benign, the advantage of (non recursive) VSFE methods, and thus VAO, is that their complexity scales on the NLC bandwidth only through $N$, as a result of an increasing the channel memory.

An accurate study on the performance/complexity trade-off between VAO and DBP requires a detailed description of the implementation of both algorithms, and is beyond the scope of this paper.

\section{Conclusion}
\label{sec:conclusion}
In this work, a novel nonlinearity compensation scheme combining mid-link optical phase conjugation (OPC) and frequency-domain Volterra equalization, was proposed. The new scheme, referred to  as Volterra-assisted OPC (VAO) was designed to overcome the limitations of both OPC and Volterra equalization in practical transmission scenarios, whilst preserving the complexity of a conventional VSFE. VAO was demonstrated to effectively recover the NLC capabilities of OPC over a large optical bandwidth, in EDFA-amplified links. The addition of mid-link OPC was shown to significantly enhance the NLC performance of a conventional VSFE. Up to 10 dB higher NLC suppression was demonstrated for VAO and 4.2 dB enhancement in the received SNR in a 1000 km link compared to either OPC or VSFE. 
As VAO dramatically outperforms Volterra-only equalisation for large NLC bandwidths, it offers in such a scenario a attractive trade-off between NLC effectiveness and computational complexity.

\appendix[Derivation of the third-order Volterra series kernel in the presence of mid-link OPC] 
\label{appendix}
We begin the derivation of the third-order Volterra kernel from the Manakov equation in the frequency domain for a dual polarization signal. Only the $X$ polarization solution will be derived here, as the solution for the $Y$ polarization is simply obtained interchanging the subscripts $X$ and $Y$. For the signal on the $X$ polarization, the Manakov equation in frequency domain reads:
\begin{equation}
\frac{\partial {A}_{\scaleto{X}{4pt}}}{\partial z}=j\frac{\beta_2}{2}\omega^2{A}_{\scaleto{X}{4pt}}-\frac{\alpha(z)}{2}{A}_{\scaleto{X}{4pt}}+j\frac{8}{9}\gamma\mathcal{F}\{(\tilde{A}_{\scaleto{X}{4pt}} \tilde{A}_{\scaleto{X}{4pt}}^*+\tilde{A}_{\scaleto{Y}{4pt}} \tilde{A}_{\scaleto{Y}{4pt}}^*)\tilde{A}_{\scaleto{X}{4pt}}\},
\label{eq:Manakov_x}
\end{equation}
where $A_{\scaleto{X}{4pt}}(\omega,z)$ is the Fourier transform of the optical field over polarization $X$ at distance $z$, whereas  
$\tilde{A}_{\scaleto{X}{4pt}}(t,z)\triangleq \mathcal{F}^{-1}\{ A_{\scaleto{X}{4pt}}\}$ is its time-domain counterpart. 
The equivalence between the Volterra series and the regular perturbation expansion terms shown in \cite{Vannucci2002} allows us to use the latter to derive the third-order Volterra kernel. Following the regular perturbation approach for the solution of \eqref{eq:Manakov_x} \cite{Vannucci2002, Johannisson2013}, the zeroth-order perturbation solution, corresponding to the first term of the Volterra series, is given by 

\begin{equation}
{A}_{1\scaleto{X}{4pt}}(\omega,N_sL_s) = {A}_{\scaleto{X}{4pt}}(\omega,0)e^{\left(-\frac{\alpha(z)}{2}+i\frac{\beta_2}{2}\omega^2\right)N_sL_s},
\label{eq:zeroth_order}
\end{equation}
where ${A}_{\scaleto{X}{4pt}}(\omega,0)$ is the Fourier transform of the transmitted optical field, $N_s$ is the number of spans and $L_s$ is the span length. The zeroth order solution is recursively used to derive the first-order perturbation term of the solution of Eq.~\eqref{eq:Manakov_x}, which corresponds to its third-order Volterra series expansion, and it is given by
\begin{equation}
\begin{split}
&{A}_{3\scaleto{X}{4pt}}(\omega,N_s L_s)= j\frac{8}{9}\gamma 
e^{\left(i\frac{\beta_{2}}{2}\omega^2-\frac{\alpha(z)}{2}\right)N_{s}L_{s}}\\
&\times\int_0^{N_sL_s} e^{\int_{0}^{z^{\prime}} (-i\frac{\beta_{2}}{2}\omega^2+\frac{\alpha}{2})dz^{\prime\prime}}D(\omega, z^{\prime})dz^{\prime},
\end{split}
\label{eq:first_order}
\end{equation}
where the term $D(\omega,z)$  comes from substituting Eq.~\eqref{eq:zeroth_order} into the nonlinear term in the left-hand side of Eq.~\eqref{eq:Manakov_x}, and it is given by
\begin{equation}
\begin{split}
&D(\omega,z)=-\iint \left[{A}_{1\scaleto{X}{4pt}}^{*}(\omega_1,z) {A}_{1\scaleto{X}{4pt}}(\omega_2,z){A}_{1\scaleto{X}{4pt}}(\omega-\omega_2+\omega_1,z) \right.\\ 
&+\left. {A}_{1{\scaleto{Y}{4pt}}}^{*}(\omega_1,z){A}_{1{\scaleto{Y}{4pt}}}(\omega_2,z){A}_{1\scaleto{X}{4pt}}(\omega-\omega_2+\omega_1,z) \right] d\omega_2 d\omega_1,
\end{split}
\label{eq:Dz}
\end{equation}
%
Substituting Eq.~\eqref{eq:Dz} into Eq.~\eqref{eq:first_order} we obtain
\begin{equation}
\begin{split}
&{A}_{3\scaleto{X}{4pt}}(\omega,N_sL_s)=j\frac{8}{9}\gamma 
e^{\left(i\frac{\beta_{2}}{2}\omega^2-\frac{\alpha(z)}{2} \right)N_{s}L_{s}}\\
&\times\iint  \left[\mathcal{S}_{\scaleto{XXX}{4pt}}(\omega,\omega_1,\omega_2)+\mathcal{S}_{\scaleto{YYX}{4pt}}(\omega,\omega_1,\omega_2)\right]\\
&\times F\left(\omega,\omega_1,\omega_2\right)\Xi\left(N_s,\omega,\omega_1,\omega_2\right)d\omega_2d\omega_1.
\end{split}
\label{eq:first_order_multispan}
\end{equation} 
where the signal kernels $\mathcal{S}_{\scaleto{XXX}{4pt}}(\omega,\omega_1,\omega_2)$ and $\mathcal{S}_{\scaleto{YYX}{4pt}}(\omega,\omega_1,\omega_2)$, the phased-array $\Xi\left(N_s,\omega,\omega_1,\omega_2\right)$, and $\Delta\Omega$ were previously defined in Sec.~\ref{sec:VAO}.
Without in-line OPC, the four-wave mixing efficiency $F\left(\omega,\omega_1,\omega_2\right)$ and the \textit{phased array} factor $\Xi\left(N_s,\omega,\omega_1,\omega_2\right)$ are defined in \eqref{eq:F1span} and \eqref{eq:pharray}, respectively.
  
To study the effect of an OPC module placed at the  mid-link point, i.e. after $\frac{N_s}{2}$ fibre spans, the solution of \eqref{eq:first_order} is conveniently split into 2 contributions: one before and one after the OPC module. Rewriting Eq.~\eqref{eq:first_order} we have: 

\begin{equation}
\begin{split}
{A}_{3\scaleto{X}{4pt}}(\omega,N_s L_s) = &i\frac{8}{9}\gamma 
e^{\left(i\frac{\beta_{2}}{2}\omega^2\right)N_{s}L_{s}}P(N_sL_s)\\
&\left[\int_0^{\frac{N_s}{2}L_s} P(-z^{\prime}) e^{\int_{0}^{z^{\prime}} -i\frac{\beta_{2}}{2}\omega^2dz^{\prime\prime}}D(z^{\prime})
 dz^{\prime}\right. \\ 
&\left.+\int_{{\frac{N_s}{2}L_s}}^{N_s L_s} P(-z^{\prime}) e^{\int_{0}^{z^{\prime}}-i\frac{\beta_{2}}{2}\omega^2dz^{\prime\prime}}
D(z^{\prime})dz^{\prime}
\right].
\end{split}
\label{eq:fwd+opc}
\end{equation}
where $P(z)$ is the power profile at distance $z$ given by 
\begin{equation}
P(z)=e^{-\int_{0}^{z}\alpha(z^{\prime})z^{\prime}}.
\label{eq:pow_profile}
\end{equation}
Both integrals in Eq.~\eqref{eq:fwd+opc} represent propagation up to specific part of the transmission link. The first one corresponds to propagation from the transmitter up the mid-link point, and the second one corresponds to propagation from the mid-link point to the end the last fiber span. Using the first-order solution for the first half of the link Eq.~\eqref{eq:fwd+opc} can be rewritten as follows:  
\begin{equation}
\begin{split}
{A}_{3\scaleto{X}{4pt}}(\omega,N_s L_s)=&{A}_{3\scaleto{X}{4pt}}\left(\omega,\frac{N_s L_s}{2}\right)P\left(\frac{N_sL_s}{2}\right)e^{i\frac{\beta_2}{2}\omega^2\frac{N_s L_s}{2}}\\ 
&+i\frac{8}{9}\gamma P(N_sL_s)e^{i\frac{\beta_2}{2}\omega^2 N_s L_s} \\
&\times\int_{{\frac{N_s}{2}L_s}}^{N_s L_s} e^{\int_{0}^{z^{\prime}}\left(-i\frac{\beta_{2}}{2}\omega^2+\frac{\alpha}{2}\right)dz^{\prime\prime}}D(z^{\prime})dz^{\prime}
\end{split}
\label{eq:1st_order_sol_opc}
\end{equation}
When an OPC module is used, both zeroth and first-order perturbation terms undergo conjugation in time-domain at $z=\frac{N_s L_s}{2}$, leading to:
\begin{equation}
\begin{split}
{A}_{3\scaleto{X}{4pt}}(\omega,N_s L_s)=&\hat{A}_{3\scaleto{X}{4pt}}\left(\omega,\frac{N_s L_s}{2}\right)P\left(\frac{N_sL_s}{2}\right)e^{i\frac{\beta_2}{2}\omega^2\frac{N_s L_s}{2}}\\ 
&+i\frac{8}{9}\gamma P(N_sL_s)e^{i\frac{\beta_2}{2}\omega^2 N_s L_s} \\
&\times\int_{{\frac{N_s}{2}L_s}}^{N_s L_s} e^{\int_{0}^{z^{\prime}}\left(-i\frac{\beta_{2}}{2}\omega^2+\frac{\alpha}{2}\right)dz^{\prime\prime}}D(z^{\prime})dz^{\prime}.
\end{split}
\end{equation}
where the conjugation of the first-order perturbation term is given by:
\begin{equation}
\begin{split}
\hat{A}_{3\scaleto{X}{4pt}}\left(\omega,\frac{N_s L_s}{2}\right)&\triangleq \mathcal{F}\left\{\tilde{A}^{*}_{3\scaleto{X}{4pt}}\left(t,\frac{N_s L_s}{2}\right)\right\}\\
&={A}^{*}_{3\scaleto{X}{4pt}}\left(-\omega,\frac{N_s L_s}{2}\right).
\end{split}
\end{equation}

The term corresponding to the first half of the transmission link then becomes:
\begin{equation}
\begin{split}
&\hat{A}_{3\scaleto{X}{4pt}}\left(\omega,\frac{N_s L_s}{2}\right)P\left(\frac{N_sL_s}{2}\right)e^{i\frac{\beta_2}{2}\omega^2\frac{N_s L_s}{2}} \\ 
&=-i\gamma\frac{8}{9}P(N_sL_s)\iint K(\omega,\omega_1,\omega_2)\Lambda^*\left(\frac{N_s}{2},\omega,\omega_1,\omega_2\right) d\omega_2 d\omega_1,
\end{split}
\end{equation}
where 
\begin{equation}
\begin{split}
\Lambda\left(\frac{N_s}{2},\omega,\omega_1,\omega_2\right)\triangleq & \sum_{n=1}^{\frac{N_s}{2}}\left[\frac{P(nL_s)e^{i\beta_2\Delta\Omega nL_s}}{\alpha(nL_s)+i\beta_2\Delta\Omega}\right. \\
&\left. -\frac{P((n-1)L_s)e^{i\beta_2\Delta\Omega (n-1)L_s}}{\alpha((n-1)L_s)+i\beta_2\Delta\Omega} \right],
\end{split}
\end{equation}
and 
\begin{equation}
K(\omega,\omega_1,\omega_2)=\mathcal{S}_{\scaleto{XXX}{4pt}}^*(-\omega,-\omega_1,-\omega_2)+\mathcal{S}_{\scaleto{YYX}{4pt}}^*(-\omega,-\omega_1,-\omega_2).
\end{equation}

Beyond the conjugation point $z\geq \frac{N_s L_s}{2}$, the zeroth order perturbative solution is 
\begin{equation}
{A}_{1\scaleto{X}{4pt}}(\omega,z)=A_{\scaleto{X}{4pt}}^{*}(-\omega,0)P(z)e^{i\frac{\beta_2}{2}\omega^2(z-N_s L_s)}.
\label{eq:zeroth_order_OPC}
\end{equation}

Substituting \eqref{eq:zeroth_order_OPC} into $D(z)$ we obtain:
\begin{equation}
\begin{split}
D(z) = &-P^3(z)e^{i\frac{\beta_2}{2}\omega^2(z-N_s L_s)}\\ 
& \iint A^*_{\scaleto{X}{4pt}}(-\omega_2,0)A_{\scaleto{X}{4pt}}(-\omega_1,0)A_{\scaleto{X}{4pt}}^*(\omega_2-\omega_1-\omega,0) \\ 
& e^{i\beta_2\Delta\Omega(z-N_s L_s)}d\omega_2 d\omega_1.
\end{split}
\end{equation}

Hence, the second term in Eq.~\eqref{eq:1st_order_sol_opc} becomes
\begin{equation}
\begin{split}
P(N_sL_s)e^{i\frac{\beta_2}{2}\omega^2 N_s L_s} \int_{{\frac{N_s}{2}L_s}}^{N_s L_s} P^{-1}(z^{\prime})e^{-i\frac{\beta_{2}}{2}\omega^2 z^{\prime}}
i\frac{8}{9} \gamma D(z^{\prime})dz^{\prime}\\
=i\gamma\frac{8}{9}P(N_sL_s)\iint K(\omega,\omega_1,\omega_2)\psi\left(\frac{N_s}{2},\omega,\omega_1,\omega_2\right) d\omega_2 d\omega_1
\end{split}
\end{equation}
where 

\begin{equation}
\begin{split}
\Psi\left(\frac{N_s}{2},\omega,\omega_1,\omega_2\right)\triangleq & e^{-i\beta_2\Delta\Omega N_s L_s}\\ 
\times\sum_{n=\frac{N_s}{2}+1}^{N_s}\left[\frac{P(nL_s)e^{i\beta_2\Delta\Omega nL_s}}{\alpha(nL_s)+i\beta_2\Delta\Omega}\right. 
& \left.-\frac{P((n-1)L_s)e^{i\beta_2\Delta\Omega (n-1)L_s}}{\alpha((n-1)L_s)+i\beta_2\Delta\Omega} \right].
\end{split}
\end{equation}

Finally, we obtain:
\begin{equation}
\begin{split}
&A_{3\scaleto{X}{4pt}}(\omega, N_sL_s)=i\gamma\frac{8}{9}P(N_sL_s)\iint K(\omega,\omega_1,\omega_2)\\
&\times\left[\Psi\left(\frac{N_s}{2},\omega,\omega_1,\omega_2\right)-\Lambda^*\left(\frac{N_s}{2},\omega,\omega_1,\omega_2\right)\right] d\omega_2 d\omega_1
\end{split}
\end{equation}

where the nonlinear kernel in the presence of OPC is given by: 
\begin{equation}
\begin{split}
\Gamma(N_s,\omega,\omega_1,\omega_2)\triangleq\Psi\left(\frac{N_s}{2},\omega,\omega_1,\omega_2\right)-\Psi^*\left(\frac{N_s}{2},\omega,\omega_1,\omega_2\right).
\end{split}
\label{eq:OPC_kernel}
\end{equation}

In order to suppress the first-order solution for all $\omega$ one condition is having $\Gamma(N_s,\omega,\omega_1,\omega_2)=0 \;\; \forall \;\; \omega, \omega_1, \omega_2$. This is verified if the following conditions are jointly verified  

\begin{eqnarray}
    P((N_s-n)L_s^-)=P((n-1)L_s^+), \\
    P((N_s-n-1)L_s^+)=P(nL_s^-),\\
  \alpha((N_s-n)L_s^-)=-\alpha((n-1)L_s^+),\\
  \alpha((N_s-n-1)L_s^+)=-\alpha(nL_s^-).
\end{eqnarray}

In the case where $\alpha=0$ (lossless fiber), all the terms in the sum in \eqref{eq:OPC_kernel} are the same and, thus, $\Gamma(N_s,\omega,\omega_1,\omega_2)=0$. If $\alpha(z)=\alpha$ everywhere except in $z=nL_s$ for $n=0,1,2,...$ where $\alpha(z)=\delta(z-nL_s)$ (EDFA amplification) we have

\begin{align}
    P((N_s-n)L_s^-)&=P(nL_s^-),\\
    P((N_s-n-1)L_s^+)&=P((n-1)L_s^+),\\ 
    \begin{split}
      \alpha((N_s-n)L_s^-)&=\alpha((n-1)L_s^+) \\=\alpha((N_s-n-1)L_s^+)
  &=\alpha(nL_s^-), \;\; \text{for} \;\; n=1,2,...,\frac{N_s}{2}
    \end{split}
\end{align}
and the OPC kernel becomes:
\begin{equation}
\Gamma(N_s,\omega,\omega_1,\omega_2)=\Xi^*\left(\frac{N_s}{2},\omega,\omega_1,\omega_2\right) G(\omega,\omega_1,\omega_2)
\end{equation}
where 
\begin{align}
\begin{split}
G(\omega,\omega_1,\omega_2)\triangleq\frac{(e^{-i\beta_2\Delta\Omega L_s}e^{-\alpha L_s}-1)(\alpha-i\beta_2\Delta\Omega)}{\alpha^2+\beta_2^2\Delta\Omega^2}\\
+\frac{(e^{-i\beta_2\Delta\Omega L_s}-e^{-\alpha L_s})(\alpha+i\beta_2\Delta\Omega)}{\alpha^2+\beta_2^2\Delta\Omega^2}
\end{split}
\label{eq:OPC_char_kernel}
\end{align}
is the characteristic kernel of the optical channel in the presence of mid-link OPC. 
The Volterra series terms discussed above corresponds to the case of signal propagation in the optical fiber. A Volterra equalizer instead is compute such terms for the \emph{reverse propagation} case corresponding to integrating \eqref{eq:Manakov_x} in the \emph{backward} propagation direction. From standard calculus operations one can obtain the Volterra equalizer terms (for the case with OPC) as 

\begin{align}
\begin{split}
{A}_{1\scaleto{X}{4pt}}(\omega,0)&={A}_{\scaleto{X}{4pt}}(\omega,N_s L_s)e^{-i\frac{\beta_2}{2}\omega^2 N_sL_s}\\
{A}_{3\scaleto{X}{4pt}}(\omega,0)&=-j\gamma\frac{8}{9}\iint K^{\prime}(\omega,\omega_1,\omega_2)\\
&\times\Xi^{*}\left(-\frac{N_s}{2},\omega,\omega_1,\omega_2\right)G^{\prime}(\omega,\omega_1,\omega_2) d\omega_2 d\omega_1
\end{split}
\label{eq:Volterra_reverse}
\end{align}

where 

\begin{align}
\begin{split}
F^{\prime}(\omega,\omega_1,\omega_2)&=\frac{1-e^{\alpha L_s} e^{-j\beta_2\Delta\Omega L_s}}{j\beta_2\Delta\Omega-\alpha},\\
G^{\prime}(\omega,\omega_1,\omega_2)&=\frac{(e^{j\beta_2\Delta\Omega L_s}e^{\alpha L_s}-1)(\alpha-j\beta_2\Delta\Omega)}{\alpha^2+\beta_2^2\Delta\Omega^2}\\
+&\frac{(e^{j\beta_2\Delta\Omega L_s}-e^{\alpha L_s})(\alpha+i\beta_2\Delta\Omega)}{\alpha^2+\beta_2^2\Delta\Omega^2}.
\end{split}
\label{eq:Volterra_reverse_terms}
\end{align}

are the \emph{inverse} Volterra kernels for backward propagation and $K^{\prime}(\omega,\omega_1,\omega_2)$ is defined as in \eqref{eq:SignalKernels} but where the transmitted signal spectrum $S(\omega)$ is replaced by the received signal spectrum.

\section*{Acknowledgment}
The authors would like to acknowledge Dr. D. Lavery (UCL) for insightful discussions on the computational complexity of NLC schemes, N. Shevchenko and D. Semrau (UCL) for discussions regarding the derivation of the Volterra series expansion for systems using mid-link OPC.
\bibliographystyle{IEEEtran}
\bibliography{IEEEabrv,Biblio}

\begin{thebibliography}{10}
\providecommand{\url}[1]{#1}
\csname url@samestyle\endcsname
\providecommand{\newblock}{\relax}
\providecommand{\bibinfo}[2]{#2}
\providecommand{\BIBentrySTDinterwordspacing}{\spaceskip=0pt\relax}
\providecommand{\BIBentryALTinterwordstretchfactor}{4}
\providecommand{\BIBentryALTinterwordspacing}{\spaceskip=\fontdimen2\font plus
\BIBentryALTinterwordstretchfactor\fontdimen3\font minus
  \fontdimen4\font\relax}
\providecommand{\BIBforeignlanguage}[2]{{%
\expandafter\ifx\csname l@#1\endcsname\relax
\typeout{** WARNING: IEEEtran.bst: No hyphenation pattern has been}%
\typeout{** loaded for the language `#1'. Using the pattern for}%
\typeout{** the default language instead.}%
\else
\language=\csname l@#1\endcsname
\fi
#2}}
\providecommand{\BIBdecl}{\relax}
\BIBdecl

\bibitem{Essiambre2010}
R.~J. Essiambre, G.~Kramer, P.~J. Winzer, G.~J. Foschini, and B.~Goebel,
  ``{Capacity Limits of Optical Fiber Networks},'' \emph{Journal of Lightwave
  Technology}, vol.~28, no.~4, pp. 662--701, Feb 2010.

\bibitem{Agrell2016}
E.~Agrell, M.~Karlsson, A.~R. Chraplyvy, D.~J. Richardson, P.~M. Krummrich,
  P.~Winzer, K.~Roberts, J.~K. Fischer, S.~J. Savory, B.~J. Eggleton,
  M.~Secondini, F.~R. Kschischang, A.~Lord, J.~Prat, I.~Tomkos, J.~E. Bowers,
  S.~Srinivasan, M.~Brandt-Pearce, and N.~Gisin, ``{Roadmap of optical
  communications},'' \emph{Journal of Optics}, vol.~18, no.~6, p. 063002, 2016.

\bibitem{Li2008}
X.~Li, X.~Chen, G.~Goldfarb, E.~Mateo, I.~Kim, F.~Yaman, and G.~Li,
  ``{Electronic post-compensation of WDM transmission impairments using
  coherent detection and digital signal processing},'' \emph{Opt. Express},
  vol.~16, no.~2, pp. 880--888, Jan 2008.

\bibitem{Ip2008}
E.~Ip and J.~M. Kahn, ``{Compensation of Dispersion and Nonlinear Impairments
  Using Digital Backpropagation},'' \emph{Journal of Lightwave Technology},
  vol.~26, no.~20, pp. 3416--3425, Oct 2008.

\bibitem{Mateo2010}
E.~F. Mateo, F.~Yaman, and G.~Li, ``{Efficient compensation of inter-channel
  nonlinear effects via digital backward propagation in WDM optical
  transmission},'' \emph{Opt. Express}, vol.~18, no.~14, pp. 15\,144--15\,154,
  Jul 2010.

\bibitem{Liga2014}
G.~Liga, T.~Xu, A.~Alvarado, R.~I. Killey, and P.~Bayvel, ``{On the performance
  of multichannel digital backpropagation in high-capacity long-haul optical
  transmission},'' \emph{Opt. Express}, vol.~22, no.~24, pp. 30\,053--30\,062,
  Dec 2014.

\bibitem{Peddanarappagari1997}
K.~V. Peddanarappagari and M.~Brandt-Pearce, ``{Volterra series transfer
  function of single-mode fibers},'' \emph{Journal of Lightwave Technology},
  vol.~15, no.~12, pp. 2232--2241, Dec 1997.

\bibitem{Gao2010}
Y.~Gao, F.~Zhang, L.~Dou, Z.~Chen, and A.~Xu, ``{Intra-channel nonlinearities
  mitigation in pseudo-linear coherent QPSK transmission systems via nonlinear
  electrical equalizer},'' \emph{Optics Communications}, vol. 282, no.~12, pp.
  2421 -- 2425, 2009.

\bibitem{Pan2011}
Z.~Pan, B.~Chatelain, M.~Chagnon, and D.~V. Plant, ``{Volterra filtering for
  nonlinearity impairment mitigation in DP-16QAM and DP-QPSK fiber optic
  communication systems},'' in \emph{2011 Optical Fiber Communication
  Conference and Exposition and the National Fiber Optic Engineers Conference},
  no.~2, 2011, p. JThA40.

\bibitem{Guiomar2011}
F.~P. Guiomar, J.~D. Reis, A.~L. Teixeira, and A.~N. Pinto, ``{Digital
  postcompensation using Volterra series transfer function},'' \emph{IEEE
  Photonics Technology Letters}, vol.~23, no.~19, pp. 1412--1414, Oct 2011.

\bibitem{Guiomar2012}
------, ``{Mitigation of intra-channel nonlinearities using a frequency-domain
  Volterra series equalizer},'' \emph{Opt. Express}, vol.~20, no.~2, pp.
  1360--1369, Jan 2012.

\bibitem{Guiomar2013}
F.~P. Guiomar and A.~N. Pinto, ``{Simplified Volterra Series Nonlinear
  Equalizer for Polarization-Multiplexed Coherent Optical Systems},''
  \emph{Journal of Lightwave Technology}, vol.~31, no.~23, pp. 3879--3891, Dec
  2013.

\bibitem{Cartledge2017}
J.~C. Cartledge, F.~P. Guiomar, F.~R. Kschischang, G.~Liga, and M.~P. Yankov,
  ``Digital signal processing for fiber nonlinearities,'' \emph{Opt. Express},
  vol.~25, no.~3, pp. 1916--1936, Feb 2017.

\bibitem{Liu2012}
L.~Liu, L.~Li, Y.~Huang, K.~Cui, Q.~Xiong, F.~N. Hauske, C.~Xie, and Y.~Cai,
  ``{Intrachannel nonlinearity compensation by inverse Volterra series transfer
  function},'' \emph{Journal of Lightwave Technology}, vol.~30, no.~3, pp.
  310--316, 2012.

\bibitem{Bakhshali2016}
A.~Bakhshali, W.~Y. Chan, J.~C. Cartledge, M.~O’Sullivan, C.~Laperle,
  A.~Borowiec, and K.~Roberts, ``{Frequency-domain Volterra-based equalization
  structures for efficient mitigation of intrachannel Kerr nonlinearities},''
  \emph{Journal of Lightwave Technology}, vol.~34, no.~8, pp. 1770--1777, April
  2016.

\bibitem{Yariv79}
A.~Yariv, D.~Fekete, and D.~M. Pepper, ``{Compensation for channel dispersion
  by nonlinear optical phase conjugation},'' \emph{Opt. Lett.}, vol.~4, no.~2,
  pp. 52--54, Feb 1979.

\bibitem{Fisher83}
R.~A. Fisher, B.~R. Suydam, and D.~Yevick, ``{Optical phase conjugation for
  time-domain undoing of dispersive self-phase-modulation effects},''
  \emph{Opt. Lett.}, vol.~8, no.~12, pp. 611--613, Dec 1983.

\bibitem{Watanabe96}
S.~Watanabe and M.~Shirasaki, ``{Exact compensation for both chromatic
  dispersion and Kerr effect in a transmission fiber using optical phase
  conjugation},'' \emph{Journal of Lightwave Technology}, vol.~14, no.~3, pp.
  243--248, Mar 1996.

\bibitem{Lorattanasane97}
C.~Lorattanasane and K.~Kikuchi, ``{Design theory of long-distance optical
  transmission systems using midway optical phase conjugation},'' \emph{Journal
  of Lightwave Technology}, vol.~15, no.~6, pp. 948--955, Jun 1997.

\bibitem{XLiu2012}
X.~Liu, A.~Chraplyvy, P.~Winzer, R.~Tkach, and S.~Chandrasekhar,
  ``{Phase-conjugated twin waves for communication beyond the Kerr nonlinearity
  limit},'' \emph{Nature Photonics}, vol.~7, no.~7, p. 560, 2013.

\bibitem{Umeki2016}
T.~Umeki, T.~Kazama, A.~Sano, K.~Shibahara, K.~Suzuki, M.~Abe, H.~Takenouchi,
  and Y.~Miyamoto, ``{Simultaneous nonlinearity mitigation in 92 $\times$
  180-Gbit/s PDM-16QAM transmission over 3840 km using PPLN-based
  guard-band-less optical phase conjugation},'' \emph{Opt. Express}, vol.~24,
  no.~15, pp. 16\,945--16\,951, Jul 2016.

\bibitem{Rosa2015}
P.~Rosa, S.~T. Le, G.~Rizzelli, M.~Tan, and J.~D. Ania-Casta{\~{n}}\'{o}n,
  ``{Signal power asymmetry optimisation for optical phase conjugation using
  Raman amplification},'' \emph{Opt. Express}, vol.~23, no.~25, pp.
  31\,772--31\,778, Dec 2015.

\bibitem{Ellis2014}
A.~D. Ellis, M.~Tan, M.~A. Iqbal, M.~A.~Z. Al-Khateeb, V.~Gordienko, G.~S.
  Mondaca, S.~Fabbri, M.~F.~C. Stephens, M.~E. McCarthy, A.~Perentos, I.~D.
  Phillips, D.~Lavery, G.~Liga, R.~Maher, P.~Harper, N.~Doran, S.~K. Turitsyn,
  S.~Sygletos, and P.~Bayvel, ``{4 Tb/s Transmission Reach Enhancement Using 10
  x 400 Gb/s Super-Channels and Polarization Insensitive Dual Band Optical
  Phase Conjugation},'' \emph{Journal of Lightwave Technology}, vol.~34, no.~8,
  pp. 1717--1723, Apr 2016.

\bibitem{Saavedra2018}
G.~Saavedra, Y.~Sun, K.~R.~H. Bottrill, L.~Galdino, F.~Parmigiani, Z.~Liu,
  D.~J. Richardson, P.~Petropoulos, R.~I. Killey, and P.~Bayvel, ``{Optical
  Phase Conjugation in Installed Optical Networks},'' in \emph{Optical Fiber
  Communication Conference}.\hskip 1em plus 0.5em minus 0.4em\relax Optical
  Society of America, 2018, p. W3E.2.

\bibitem{Liga2018}
G.~Liga, G.~Saavedra, and P.~Bayvel, ``{Combining Optical Phase Conjugation and
  Volterra Equalisation: a Novel Nonlinearity Compensation Scheme},'' in
  \emph{ECOC 2018; 44th European Conference on Optical Communication, Rome,
  Italy}.\hskip 1em plus 0.5em minus 0.4em\relax IEEE, 2018, p. Tu1F.4.

\bibitem{Cartledge2016}
J.~C. Cartledge, A.~Kashi, A.~D. Ellis, M.~Mccarthy, A.~Shiner, M.~Reimer, and
  A.~Borowiec, ``{Signal Processing Techniques for Reducing the Impact of Fiber
  Nonlinearities on System Performance},'' no.~1, pp. 16--18.

\bibitem{Vannucci2002}
A.~Vannucci, P.~Serena, and A.~Bononi, ``{The RP method: a new tool for the
  iterative solution of the nonlinear Schrodinger equation},'' \emph{Journal of
  Lightwave Technology}, vol.~20, no.~7, pp. 1102--1112, Jul 2002.

\bibitem{Taghavi2006}
M.~Taghavi, G.~Papen, and P.~Siegel, ``{On the Multiuser Capacity of WDM in a
  Nonlinear Optical Fiber: Coherent Communication},'' \emph{IEEE Transactions
  on Information Theory}, vol.~52, no.~11, pp. 5008--5022, nov 2006.

\bibitem{Pechenkin2010}
V.~Pechenkin and I.~J. Fair, ``{Analysis of four-wave mixing suppression in
  fiber-optic OFDM transmission systems with an optical phase conjugation
  module},'' \emph{IEEE/OSA Journal of Optical Communications and Networking},
  vol.~2, no.~9, pp. 701--710, September 2010.

\bibitem{Al-Khateeb2018}
M.~A.~Z. Al-Khateeb, M.~A. Iqbal, M.~Tan, A.~Ali, M.~McCarthy, P.~Harper, and
  A.~D. Ellis, ``{Analysis of the nonlinear Kerr effects in optical
  transmission systems that deploy optical phase conjugation},'' \emph{Opt.
  Express}, vol.~26, no.~3, pp. 3145--3160, Feb 2018.

\bibitem{Al-Khateeb:18}
M.~A.~Z. Al-Khateeb, M.~E. McCarthy, C.~S\'{a}nchez, and A.~D. Ellis,
  ``{Nonlinearity compensation using optical phase conjugation deployed in
  discretely amplified transmission systems},'' \emph{Opt. Express}, vol.~26,
  no.~18, pp. 23\,945--23\,959, Sep 2018.

\bibitem{Xu2002}
B.~Xu and M.~Brandt-Pearce, ``{Modified Volterra series transfer function
  method},'' \emph{Photonics Technology Letters, IEEE}, vol.~14, no.~1, pp.
  47--49, 2002.

\bibitem{Weidenfeld2010}
R.~Weidenfeld, M.~Nazarathy, R.~Noe, and I.~Shpantzer, ``{Volterra Nonlinear
  Compensation of 100G Coherent OFDM with Baud-Rate ADC, Tolerable Complexity
  and Low Intra-Channel FWM/XPM Error Propagation},'' in \emph{Optical Fiber
  Communication Conference}, 2010, p. OTuE3.

\bibitem{Shulkind2013}
G.~Shulkind and M.~Nazarathy, ``{Nonlinear Digital Back Propagation compensator
  for coherent optical OFDM based on factorizing the Volterra Series Transfer
  Function},'' \emph{Optics Express}, vol.~21, no.~11, p. 13145, 2013.

\bibitem{Secondini2016}
M.~Secondini, S.~Rommel, G.~Meloni, F.~Fresi, E.~Forestieri, and L.~Pot{\`{i}},
  ``{Single-step digital backpropagation for nonlinearity mitigation},''
  vol.~31, no.~3, pp. 1--10, 2015.

\bibitem{Bosco2000}
G.~Bosco, A.~Carena, V.~Curri, R.~Gaudino, P.~Poggiolini, and S.~Benedetto,
  ``{Suppression of Spurious Tones Induced by the Split-Step Method in Fiber
  Systems Simulation},'' \emph{IEEE Photonics Technology Letters}, vol.~12,
  no.~5, pp. 489--491, 2000.

\bibitem{Johannisson2013}
P.~Johannisson and M.~Karlsson, ``{Perturbation Analysis of Nonlinear
  Propagation in a Strongly Dispersive Optical Communication System},''
  \emph{Journal of Lightwave Technology}, vol.~31, no.~8, pp. 1273--1282, apr
  2013.

\end{thebibliography}
\end{document}